\pdfoutput=1
\documentclass[doublecolumn]{aa}

\usepackage{graphicx}
\usepackage{times,epsfig}
\usepackage{psfrag}
\usepackage{mathrsfs}
\usepackage{natbib}
\bibpunct{(}{)}{;}{a}{}{,} 
\usepackage{amsmath}
\usepackage[english]{babel}
\usepackage{longtable}
\usepackage{supertabular}
\usepackage{url}
\usepackage{bm}
\usepackage{rotating}
\usepackage{times}
\usepackage{amsmath,amssymb}
\usepackage{color}

\def\aj{AJ}
\def\araa{ARA\&A}
\def\apj{ApJ}
\def\apjl{ApJL}
\def\apjs{ApJS}
\def\ao{Appl.~Opt.}
\def\aap{A\&A}
\def\mnras{MNRAS}
\def\pasp{PASP}
\def\nat{Nature}
 
\DeclareMathAlphabet{\mathsc}{OT1}{cmr}{m}{sc} 
\def\testbx{bx}%
\DeclareRobustCommand{\ion}[2]{%
\relax\ifmmode 
\ifx\testbx\f@series 
{\mathbf{#1\,\mathsc{#2}}}\else 
{\mathrm{#1\,\mathsc{#2}}}\fi 
\else\textup{#1\,{\mdseries\textsc{#2}}}%
\fi} 
\newcommand{\ha} {\mbox{H$\alpha$}} 
\newcommand{\hb} {\mbox{H$\beta$}}

\newcommand{\Feii} {\ion{Fe}{ii}} 
\newcommand{\Feiii} {\ion{Fe}{iii}} 
 
\newcommand{\Caiia} {[\ion{Ca}{ii}]} 
 
\newcommand{\Caii} {\ion{Ca}{ii}}

\newcommand{\Nai} {\ion{Na}{i}} 
\newcommand{\Mgi} {\ion{Mg}{i}} 
\newcommand{\Mgia} {\ion{Mg}{i}]} 
\newcommand{\Mgii} {\ion{Mg}{ii}}

\newcommand{\Niia} {[\ion{N}{ii}]} 
\newcommand{\Oia} {[\ion{O}{i}]} 
\newcommand{\Oiia} {[\ion{O}{ii}]} 
\newcommand{\Oi} {\ion{O}{i}} 
\newcommand{\Oii} {\ion{O}{ii}} 
\newcommand{\Oiii} {\ion{O}{iii}} 
\newcommand{\Oiiia} {[\ion{O}{iii}]} 
\newcommand{\Cia} {[\ion{C}{i}]}

\newcommand{\Siia} {[\ion{S}{ii}]} 
 
\newcommand{\SiII} {\ion{Si}{ii}}

\def\sn{SN 2012aa}

\newcommand{\cbv}{\mbox{$(B\!-\!V)_0$}} 
\newcommand{\cvr}{\mbox{$(V\!-\!R_c)_0$}} 
\newcommand{\cvi}{\mbox{$(V\!-\!I_c)_0$}} 
\newcommand{\cgr}{\mbox{$(g\!-\!r)_0$}} 
\newcommand{\cri}{\mbox{$(r\!-\!i)_0$}} 
\newcommand{\crz}{\mbox{$(r\!-\!z)_0$}} 
\newcommand{\ebv}{\mbox{$E(B-V)$}} 
\newcommand{\bvri}{\mbox{$BVR_cI_c$}} 
\newcommand{\degree}{\mbox{$^\circ$}} 
\newcommand{\msun}{\mbox{M$_{\odot}$}} 
\newcommand{\zsun}{\mbox{Z$_{\odot}$}} 
 
\newcommand{\kms}{\mbox{$\rm{\,km\,s^{-1}}$}} 
\newcommand{\ergs}{\mbox{$\rm{\,erg\,s^{-1}}$}}

\newcommand{\mum}{\mbox{$\mu{\rm m}$}}

\begin{document}

\title{\sn\ $-$ a transient between Type Ibc core-collapse and superluminous
 supernovae}

\author{R. Roy\inst{1}
\and J. Sollerman\inst{1}
\and J. M. Silverman\inst{2}
\and A. Pastorello\inst{3}
\and C. Fransson\inst{1}
\and A. Drake\inst{4}
\and F. Taddia\inst{1}
\and C. Fremling\inst{1}
\and E. Kankare\inst{5}
\and B. Kumar\inst{6}
\and E. Cappellaro\inst{3}
\and S. Bose\inst{7}
\and S. Benetti\inst{3}
\and A. V. Filippenko\inst{8}
\and S. Valenti\inst{9}
\and A. Nyholm\inst{1}
\and M. Ergon\inst{1}
\and F. Sutaria\inst{6}
\and B. Kumar\inst{7}
\and S. B. Pandey\inst{7}
\and M. Nicholl\inst{5}
\and D. Garcia-$\acute{\rm A}$lvarez\inst{10,11,12}
\and L. Tomasella\inst{3}
\and E. Karamehmetoglu\inst{1}
\and K. Migotto\inst{1}
}

\institute{The Oskar Klein Centre, Department of Astronomy, Stockholm University, AlbaNova, 10691 Stockholm, Sweden.\\ (\email{rupak.roy@astro.su.se, rupakroy1980@gmail.com})
\and Department of Astronomy, University of Texas, Austin, TX 787120259, USA.
\and INAF-Osservatorio Astronomico di Padova, Vicolo dell'Osservatorio 5, 35122 Padova, Italy.
\and California Institute of Technology, 1200 E. California Blvd., CA 91225, USA.
\and Astrophysics Research Centre, School of Mathematics and Physics, Queen's University Belfast, BT7 1NN, UK.
\and Indian Institute of Astrophysics, Koramangala, Bangalore 560 034, India.
\and Aryabhatta Research Institute of Observational Sciences (ARIES), Manora Peak, Nainital, 263 129, India.
\and Department of Astronomy, University of California, Berkeley, CA 94720-3411, USA.
\and Department of Physics, University of California, Davis, CA, USA.
\and Instituto de Astrofísica de Canarias, E-38205 La Laguna, Tenerife, Spain.
\and Dpto. de Astrofísica, Universidad de La Laguna, E-38206 La Laguna, Tenerife, Spain.
\and Grantecan CALP, E-38712, Bre$\bar{\rm n}$a Baja, La Palma, Spain.
}
\date{Received / Accepted}

\abstract
{Research on supernovae (SNe) over the past decade has confirmed that there is a
 distinct class of events which are much more luminous (by $\sim2$ mag) than
 canonical core-collapse SNe (CCSNe). These events with visual peak magnitudes
 $\lesssim-21$ are called superluminous SNe (SLSNe). The mechanism for
 powering the light curves of SLSNe is still not well understood. The proposed
 scenarios are circumstellar interaction, the emergence of a magnetar after core
 collapse, or disruption of a massive star through pair production.}
{There are a few intermediate events which have luminosities between these two
 classes. They are important for constraining the nature of the progenitors of
 these two different populations as well as their environments and powering
 mechanisms. Here we study one such object, SN 2012aa.}
{We observed and analysed the evolution of the luminous Type Ic \sn. The event
 was discovered by the Lick Observatory Supernova Search in an anonymous
 ($z\approx0.08$) galaxy. The optical photometric and spectroscopic follow-up
 observations were conducted over a time span of about 120 days.} 
{With an absolute $V$-band peak of $\sim-20$ mag, the SN is an
 intermediate-luminosity transient between regular SNe~Ibc and SLSNe. \sn\ also
 exhibits an unusual secondary bump after the maximum in its light curve. For
 \sn, we interpret this as a manifestation of SN-shock interaction with the
 circumstellar medium (CSM). If we would assume a $^{56}$Ni-powered ejecta, the
 quasi-bolometric light curve requires roughly 1.3 \msun\ of $^{56}$Ni and an
 ejected mass of $\sim 14$ \msun. This would also imply a high kinetic energy of
 the explosion, $\sim5.4\times10^{51}$ ergs. On the other hand, the unusually
 broad light curve along with the secondary peak indicate the possibility of
 interaction with CSM. The third alternative is the presence of a central engine releasing spin energy that eventually powers the light curve over a long time.
 The host of \sn\ is a star-forming Sa/Sb/Sbc galaxy.}
{Although the spectral properties of \sn\ and its velocity evolution are
 comparable to those of normal SNe~Ibc, its broad light curve along with a
 large peak luminosity distinguish it from canonical CCSNe, suggesting the
 event to be an intermediate-luminosity transient between CCSNe and SLSNe at
 least in terms of peak luminosity. Comparing to other SNe, we argue that \sn\
 belongs to a subclass where CSM interaction plays a significant role in
 powering the SN, at least during the initial stages of evolution.
}

\keywords{supernovae: general --- supernovae: individual (SN 2012aa)} 

\maketitle

\section{Introduction} \label{intro}
 The study of SLSNe \citep{2012Sci...337..927G} has emerged from the development
 of untargeted transient surveys such as the Texas Supernova Search
 \citep{2006PhDT........13Q}, the Palomar Transient Factory (PTF;
 \citealt{2009PASP..121.1334R}), the Catalina Real-time Transient Survey (CRTS;
 \citealt{2009ApJ...696..870D}), and the Panoramic Survey Telescope \& Rapid
 Response System (Pan-STARRS; \citealt{2004AN....325..636H}). SLSNe are much
 more luminous than normal core-collapse SNe (CCSNe;
 \citealt{1997ARA&A..35..309F}). It was with the discovery of objects such as
 SNe 2005ap and SCP-06F6 \citep{2007ApJ...668L..99Q, 2011Natur.474..487Q}, as
 well as SN 2007bi \citep{2009Natur.462..624G} that events with peak
 luminosities $\gtrsim7\times10^{43}$erg\,s$^{-1}$ ($\lesssim-21$ absolute mag,
 over 2 mag brighter than the bulk CCSN population) became known.

 SLSNe have been classified into three groups: SLSN-I, SLSN-II and SLSN-R
 \citep{2012Sci...337..927G}. The SLSN-II are hydrogen (H) rich (e.g., SN
 2006gy, \citealt{2007ApJ...666.1116S}; CSS100217,
 \citealt{2011ApJ...735..106D}; CSS121015, \citealt{2014MNRAS.441..289B}), while
 the others are H-poor. The SLSN-R (e.g., SN 2007bi,
 \citealt{2009Natur.462..624G}) have post-maximum decline rates consistent with
 the $^{56}$Co $\rightarrow$ $^{56}$Fe radioactive decay, whereas SLSNe-I (e.g.,
 SNe 2010gx, \citealt{2010ApJ...724L..16P}; SCP-06F6) have steeper declines.
 H-poor events mostly exhibit SN~Ic-like spectral evolution
 \citep{2010ApJ...724L..16P, 2011Natur.474..487Q, 2013ApJ...770..128I}. However,
 the explosion and emission-powering mechanisms of these transients are still
 disputed. It is also not clear whether CCSNe and SLSNe are originated from
 similar or completely different progenitor channels, or if there is a link
 between these kinds of explosions. In particular, the SLSNe-R were initially
 thought to be  powered by radioactive decay, but later the energy release of a
 spin-down magnetar was proposed \citep{2013ApJ...770..128I}, as was the
 CSM-interaction scenario \citep{2015ApJ...814..108Y, 2015arXiv151000834S}. In
 this present work, although we use the notation of the initial classification
 scheme, by SLSNe-R we simply mean those Type Ic SLSNe which show shallow (or
 comparable to $^{56}$Co decay) decline after maximum brightness, while the
 events which decline faster than $^{56}$Co decay are designated as SLSNe-I.

 The basic mechanisms governing the emission of stripped-envelope CCSNe are
 relatively well known. At early times ($\lesssim5$\,d after explosion), SNe~Ibc
 are powered by a cooling shock \cite{2013ApJ...769...67P, 2015A&A...574A..60T}.
 Thereafter, radioactive heating ($^{56}$Ni $\rightarrow$ $^{56}$Co) powers the
 emission from the ejecta. Beyond the peak luminosity, the photosphere cools 
 and eventually the ejecta become optically thin. By $\sim100$\,d post explosion
 the luminosity starts to decrease linearly (in mag) with time, and it is
 believed that $^{56}$Co is the main source powering the light curve during
 these epochs (\citealt{1982ApJ...253..785A}, and references therein).

\begin{figure}
\centering
\includegraphics[width=8.5cm]{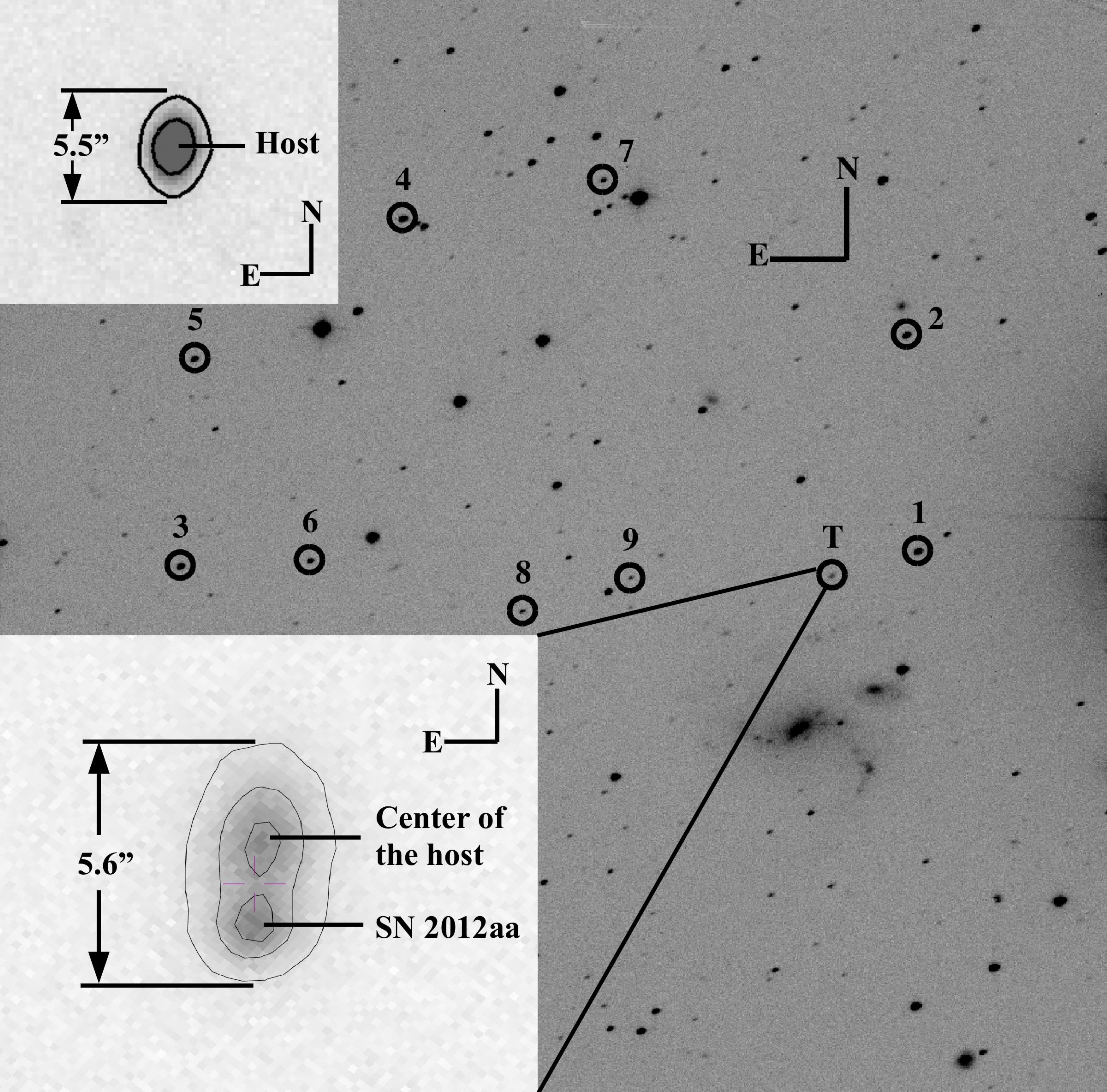}
\caption{Identification chart of SN 2012aa. The image is about $10'$ on a
 side, taken in the $R_c$ band with the 1.04\,m Sampurnanand telescope at ARIES,
 Nainital. The local standard stars are numbered. North is up and east
 is to the left. The location of the target is marked ``T'' --- as 
 observed with the 8\,m Gemini
 South telescope (bottom-left inset) on 2012 April 26,  +88\,d after
 discovery. The SN is resolved; its position and the host center are marked.
 The field was also observed with the 3\,m TNG (top-left inset) on 2013 
 April 9, long after the SN had faded away. The SN location is not 
 resolved in the TNG image.}
\label{fig:snid}
\end{figure}

 Three different models have been suggested for the explosion mechanism of
 SLSNe. A pair-instability supernova (PISN) produces a large mass of radioactive
 $^{56}$Ni, a source of enormous optical luminosity \citep{1967PhRvL..18..379B,
 1967ApJ...148..803R}. Alternatively, a similarly large luminosity can be
 produced by interaction between the SN ejecta and a dense circumstellar medium
 (CSM) surrounding the progenitor, and the radiation is released when the strong
 shocks convert the bulk kinetic energy of the SN ejecta to thermal energy
 \citep{2007ApJ...671L..17S, 2011ApJ...729L...6C, 2012ApJ...757..178G}. The
 third alternative is to invoke an engine-driven explosion: a protomagnetar is
 a compact remnant that radiates enormous power during its spin-down process
 \citep{2010ApJ...717..245K, 2010ApJ...719L.204W}. Although both magnetar
 engines and CSM interaction can reproduce the range of luminosities and shapes
 of SLSN light curves, there are very few SNe that show a PISN-like $^{56}$Co
 $\rightarrow$ $^{56}$Fe decay rate after maximum light. Only $\sim10$\% of
 SLSNe show a slowly fading tail \citep{2013Natur.502..346N,
 2015MNRAS.448.1206M}, but do not exactly follow the radioactive decay rate of
 cobalt. SLSNe 2006gy and 2007bi could be explained as PISN producing
 $\sim5$--10 \msun\ of $^{56}$Ni \citep{2007AIPC..937..412N,
 2009Natur.462..624G}, but there are also counterarguments because the rise
 times of some similar events are shorter than the PISN model predictions
 (\citealt{2011ApJ...734..102K, 2013MNRAS.428.1020M, 2013Natur.502..346N}, and
 references therein). 

 Beyond the general evolutionary picture of Type I SNe (hereafter we exclude
 SNe~Ia), many events exhibit different kinds of peculiarities. One of these,
 observed in SN 2005bf, is the appearance of a short-lived bump before the
 principal light-curve peak \citep{2005ApJ...631L.125A}. Two hypotheses have
 been proposed to explain this early short-duration bump: either a burst of
 radiation from a blob containing radioactive $^{56}$Ni that was ejected
 asymmetrically \citep{2007ApJ...666.1069M}, or a polar explosion powered by a
 relativistic jet \citep{2006ApJ...641.1039F}. Although a double-peaked light
 curve was observed in only a single SN~Ic, there are at least a handful ($\sim
 8$) Type I SLSNe that exhibit a short-duration bump before the principal broad
 peak \citep{2016MNRAS.457L..79N}. Most of them have been described using the
 magnetar model. For SN-LSQ14bdq it was proposed that the intense radiation in
 optical and ultraviolet wavelengths was due to the emergence of the shock
 driven by a high-pressure bubble produced by the central engine, a
 proto-magnetar\citep{2016ApJ...821...36K}.

 Since H-poor SLSNe and normal Type Ic events show similar spectral features
 beyond maximum light, it is worth to explore whether there is any link between
 them. In fact the gap between SNe and SLSNe has not been well explored. The
 Type Ib SN 2012au showed spectral signatures which mark the transition between
 H-poor CCSNe and SLSNe, although its peak magnitude was comparable to those
 of H-poor (stripped envelope) CCSNe \citep{2013ApJ...770L..38M}. Recently, a
 few transients such as PTF09ge, PTF09axc, PTF09djl, PTF10iam, PTF10nuj, and
 PTF11glr with peak absolute magnitudes between $-19$ and $-21$ were discovered
 \citep{2014ApJ...793...38A}. Except PTF09ge, all had H-dominated
 spectra, and except PTF10iam they showed spectral behavior like that of tidal
 disruption events (TDEs; \citealt{1988Natur.333..523R}). However,
 note that TDEs may be even more luminous than SLSNe --- for example, the ROTSE
 collaboration found an event (nicknamed Dougie) that had a peak magnitude 
 of $\sim-22.5$ \citep{2015ApJ...798...12V}. Unlike CCSNe or SLSNe, TDEs are 
 always found at the center of their host galaxies, and their spectral features
 are normally dominated by H/He emission lines, thus differing from Type Ic
 SNe. \citet{2016ApJ...819...35A} also found four SNe having rapid rise times
 ($\sim10$\,d) and peak magnitudes of $\sim-20$ along with an \ha\ signature in
 their spectra (see Sect. \ref{res:otherpossibility}).

 In this work, we discuss the evolution of a SN~Ic that shows properties
 intermediate between those of SNe~Ic and H-poor SLSNe. \sn\ (= PSN
 J14523348-0331540) was discovered on 2012 January 29.56 UT (JD 2,455,956.04) in
 a relatively distant anonymous galaxy (redshift $z = 0.083$; see Sect.
 \ref{DistExt}) during the Lick Observatory Supernova Search (LOSS;
 \citealt{2001ASPC..246..121F}) with the Katzman Automatic Imaging Telescope
 (KAIT). An image of the field of \sn\ is shown in Fig.~\ref{fig:snid}. Basic
 properties of the host galaxy are given in Table~\ref{tab:propgal}. The event
 was observed at $\sim17.7$ mag (unfiltered) at discovery, and was further
 monitored in the subsequent nights and then spectroscopically classified as a
 SN~Ic \citep{2012CBET.3015....1C} on February 2 (UT dates are used throughout
 this paper). It was not observed in the X-ray in radio domains. However,
 independent optical photometry along with spectroscopic follow-up observations
 were obtained using various telescopes around the globe. We find (Sect.
 \ref{phot}) that the event was actually discovered near its peak, but it was
 caught in its rising phase by the CRTS survey \citep{2009ApJ...696..870D}. It
 shows some spectroscopic resemblance to H-poor SLSNe, although photometrically
 the peak is about 1\,mag fainter than that of typical SLSNe. More remarkably,
 \sn\ shows an unusual secondary bump after a broad primary peak in its light
 curve (Sect. \ref{phot}).
\begin{table}
\centering
\caption{\label{tab:propgal}Properties of SN 2012aa and its host galaxy}
\vskip 0.2cm
\scriptsize
\begin{tabular}{llc}
\hline\hline
     Parameters&Value&Ref\\
     \hline
     {\bf Anonymous Host:}&&\\
     Type& Sa/Sb/Sbc&Sect. \ref{Host}\\
     \\
     Position & $\alpha_{\rm J2000} = 14^{\rm h} 52^{\rm m} 33\fs55$&Sect. \ref{obs}\\
              & $\delta_{\rm J2000} = -03\degr 31\arcmin 53\farcs82$&  \\
     \\
     Abs. magnitude& $M_{B}=-18.13$ mag&Sect. \ref{Host}\\
     \\
     Redshift& $z=0.0830\pm0.0005$ &Sect. \ref{DistExt}\\
     \\
     \\
     Distance& $D=380.2$\,Mpc&Sect. \ref{DistExt}\\
     \\
     Distance modulus& $\mu \approx 37.9$\,mag&Sect. \ref{DistExt}\\
     \\
     Projected dimension& $\sim 6$\arcsec &Sect. \ref{Host}\\
     \\
     Metallicity of the host& Z$_{\rm host}\approx0.92\pm0.34$\,\zsun\,&Sect. \ref{Host}\\
     \\
     {\bf SN 2012aa:}&&\\
     Position & $\alpha_{\rm J2000} = 14^{\rm h} 52^{\rm m} 33\fs56$&Sect. \ref{obs}\\
              & $\delta_{\rm J2000} = -03\degr 31\arcmin 55\farcs45$&  \\
     \\
      Separation from center&$\sim3$ kpc&Sect. \ref{Host}\\
     \\
     Discovery date (UT)& 29.56 January 2012&Sect. \ref{intro}\\
                   & (JD 2,455,956.04)& \\ 
     \\
     Epoch of $V$ maximum: &JD 2,455,952&Sect. \ref{phot}\\ 
     \\
     Total reddening toward SN: & \ebv\,$=0.12$\,mag&Sect. \ref{DistExt}\\ 
     \hline
\end{tabular}
\end{table}

 Here we present an optical photometric and low-resolution spectroscopic
 investigation of \sn. We also make comparisons among \sn, SLSNe, and CCSNe. The
 paper is organized as follows. Section~\ref{obs} presents a brief description
 of the observations of \sn\ along with a description of the data-reduction
 procedure. The distance and extinction are estimated in Sect.~\ref{DistExt}.
 In Sect.~\ref{phot}, we study the light-curve evolution, while the evolution
 of the colours and of the quasi-bolometric luminosity are discussed in Sect.
 \ref{ColBol}. The spectroscopic evolution is presented in Sect. \ref{spec}, and
 the properties of the host galaxy in Sect. \ref{Host}. Section~\ref{Progenitor}
 explores different physical parameters of the explosion, and a comparative
 study of this event with SLSNe and CCSNe is given in Sect. \ref{res:comp}. 
 Finally, Sect.~\ref{concl} summarizes our conclusions.

\section{Observations and data reduction} \label{obs} 
\subsection{Optical multiband photometry} \label{obs:phot} 
 The optical broadband Johnson $BV$ and Cousins $R_cI_c$ follow-up observations
 started on 2012 February 10, using the 1.04\,m Sampurnanand Telescope (ST)
 plus imaging camera\footnote{A 2048 $\times$ 2048 pixel CCD camera having $24
 \times 24$ $\mu$m pixels and a scale of $0\farcs38$ pixel$^{-1}$, mounted at
 the $f$/13 Cassegrain focus of the telescope, was used for the observations.
 The gain and readout noise of the CCD camera are 10 electrons per
 analog-to-digital unit and 5.3 electrons, respectively. We performed the
 observations in a binned mode of $2 \times 2$ pixels.} at Nainital, India, and
 were conducted at 17 different epochs between +12\,d and +116\,d after the
 discovery. The field was also observed in 3 epochs with the 1.82\,m Copernico
 Telescope (CT) plus AFOSC at the Asiago Observatory and in 2 epochs with the
 3.6\,m New Technology Telescope (NTT) plus EFOSC2 at ESO, Chile, between
 +44\,d and +93\,d after discovery. The pre-processing was done using standard
 data-reduction software IRAF\footnote{IRAF stands for Image Reduction and
 Analysis Facility, distributed by the National Optical Astronomy Observatories,
 which are operated by the Association of Universities for Research in
 Astronomy, Inc. under cooperative agreement with the National Science
 Foundation.}, and photometric measurements were performed on the coadded
 frames through point-spread-function (PSF) fitting and background-subtraction
 methods, using the stand-alone version of DAOPHOT\footnote{DAOPHOT stands for
 Dominion Astrophysical Observatory Photometry.} \citep{1987PASP...99..191S}.
 The field was calibrated in the $BVR_cI_c$ bands using
 \citet{1992AJ....104..340L} standard stars from the fields of SA107,
 PG1633+099, and PG1525+071 observed on 2014 April 27 under moderate seeing
 (full width at half-maximum intensity [FWHM] $\approx$ 2\arcsec\ in the $V$
 band) and clear sky conditions\footnote{We have used the mean values of the
 atmospheric extinction coefficients of the site \citep[namely 0.28, 0.17, 0.11,
 and 0.07 mag per unit airmass for the $B$, $V$, $R_c$, and $I_c$ bands,
 respectively, from][] {2000BASI...28..675K} with typical standard deviations
 between the transformed and the standard magnitudes of Landolt stars of
 0.04\,mag in $B$, 0.02\,mag in $V$ and $R_c$, and 0.01\,mag in $I_c$.}. Nine
 isolated nonvariable stars (marked in Fig. \ref{fig:snid}) in the field of the
 transient were used as local standards to derive the zero point for the SN
 images at each epoch. The calibrated magnitudes of these 9 secondary standards
 are given in Table \ref{tab:photstd}.
\begin{table*}
\centering
\caption{Photometry of secondary standard stars in the field of SN 2012aa.}
\vskip 0.2cm
\scriptsize
\begin{tabular}{ccccccc}
\hline\hline
 Star&$\alpha_{\rm J2000}$&$\delta_{\rm J2000}$&$B$&$V$&$R_c$&$I_c$\\
       ID& (h m s)& (\degr\,\arcmin\,\arcsec)& (mag)& (mag)& (mag)& (mag)\\
\hline
      1&14 52 29.53&$-$03 31 33.50&17.24$\pm$0.04&16.57$\pm$0.02&16.14$\pm$0.03&15.71$\pm$0.03\\
      2&14 52 30.50&$-$03 29 02.40&17.62$\pm$0.04&16.88$\pm$0.02&16.42$\pm$0.03&15.96$\pm$0.03\\
      3&14 53 03.87&$-$03 32 08.50&17.86$\pm$0.05&16.99$\pm$0.03&16.42$\pm$0.02&15.89$\pm$0.03\\
      4&14 52 54.06&$-$03 27 59.80&18.36$\pm$0.06&17.07$\pm$0.02&16.23$\pm$0.02&15.38$\pm$0.02\\
      5&14 53 03.54&$-$03 29 44.20&18.02$\pm$0.04&17.32$\pm$0.03&16.85$\pm$0.02&16.38$\pm$0.02\\
      6&14 52 57.71&$-$03 32 00.00&18.25$\pm$0.06&17.59$\pm$0.04&17.16$\pm$0.03&16.76$\pm$0.03\\
      7&14 52 44.82&$-$03 27 23.40&18.45$\pm$0.05&17.96$\pm$0.04&17.60$\pm$0.04&17.19$\pm$0.06\\
      8&14 52 47.74&$-$03 32 29.20&19.16$\pm$0.06&18.74$\pm$0.08&18.42$\pm$0.05&18.06$\pm$0.05\\
      9&14 52 42.72&$-$03 32 02.50&19.54$\pm$0.08&19.10$\pm$0.08&18.78$\pm$0.07&18.41$\pm$0.08\\
\hline
\end{tabular}
\tablefoot{Magnitude uncertainties represent the root-mean square scatter 
of the night-to-night repeatability over the entire period of SN monitoring.}
\label{tab:photstd}
\end{table*}

 The SN was not resolved in the images taken with 1--3\,m telescopes (see
 Fig. \ref{fig:snid}). The host along with the SN appeared as a point source
 (T) in all frames taken from ST and CT, while it is seen to be slightly
 elongated in the NTT images. It was resolved only in the
 spectroscopic-acquisition image obtained with the Gemini-South telescope at
 +88\,d after discovery under excellent seeing conditions (FWHM
 $\approx0\farcs58$; see the bottom-left inset of Fig. \ref{fig:snid}).
 Performing astrometry on the image, we calculate the position of the host
 center as $\alpha_{\rm center} = 14^{\rm h} 52^{\rm m} 33\fs55$,
 $\delta_{\rm center} = -03\degr 31\arcmin 53\farcs82$, while that of the SN is
 $\alpha_{\rm SN} = 14^{\rm h} 52^{\rm m} 33\fs56$, $\delta_{\rm SN} =
 -03\degr 31\arcmin 55\farcs45$. The projected dimension (major axis) of the
 galaxy is $\sim5\farcs6$, while the separation between the SN and the host
 center is $\sim1\farcs96$ in the plane of the sky. As most of the photometric
 observations have worse seeing than the separation between SN and host center,
 photometry of these images provides an estimate of the total flux of the entire
 system T (supernova plus host). 

 In order to calculate the magnitude of the transient, an estimate of the host
 magnitude is required. For this purpose, the field was again observed in the
 $BVR_cI_c$ bands after the SN had faded --- first on 2013 April 9 using the
 3.5\,m Telescopio Nazionale Galileo (TNG) plus the LRS camera, and then again
 on 2014 April 27, the night for the photometric calibration at ST. The
 $R_c$-band image taken with the TNG is shown in the top-left inset of Fig.
 \ref{fig:snid}. Aperture photometry (with an aperture diameter of
 $\sim5\farcs6$) was performed to estimate the average magnitudes of the host:
 $20.17\pm0.06$, $19.29\pm0.04$, $18.67\pm0.05$, and $18.04\pm0.04$\,mag in the
 $B$, $V$, $R_c$, and $I_c$ bands, respectively. When determining the magnitude
 of the transient in every band, the flux of the host in the corresponding band
 has been subtracted from the calibrated flux of T. The photometry of \sn\ is
 given in Table \ref{tab:photlog}. Since the host is relatively distant, we have
 calculated the $K$-correction term for every band at each epoch of observation
 using the existing spectra\footnote{For each filter system, first the
 $K$-corrections were calculated for those epochs where spectra are available by
 following the methodology by \citet{1968ApJ...154...21O}. The corresponding
 spectra were convolved using Johnson and Cousins filter response curves adopted
 by \citet{2014A&A...562A..17E}. Then the $K$-corrections for each photometric
 epoch was determined by interpolating the values calculated on the
 spectroscopic nights.} (Sect. \ref{obs:spec}). $K$-corrections for all
 photometric epochs are also reported in Table \ref{tab:photlog}.
\begin{table*}
\centering
\caption{$BVR_cI_c$ photometry of SN 2012aa\tablefootmark{a}.\label{tab:photlog}}
\vskip 0.2cm
\scriptsize
\begin{tabular}{ccccrcccccccc}
\hline\hline
UT Date&JD $-$ &Phase\tablefootmark{b}&$B$&$K_{BB}$\tablefootmark{c}&$V$&$K_{VV}$\tablefootmark{c}&$R_c$&$K_{RR}$\tablefootmark{c}&$I_c$&$K_{II}$\tablefootmark{c}&Seeing\tablefootmark{d}&Telescope\tablefootmark{e}\\
(yyyy/mm/dd)& 2,450,000& (day)& (mag)& (mag)& (mag)& (mag)& (mag)& (mag)& (mag)& (mag)&(\arcsec)\\
\hline
 2012/02/11.09    &5968.59&  +16.59& 18.87$\pm$0.07     & 0.05    &18.20$\pm$0.09& 0.08    &18.09$\pm$0.08&0.19&17.91$\pm$0.06&0.20     &2.9& ST\\
 2012/02/19.24    &5976.74&  +24.74& 19.32$\pm$0.05     & 0.04    &18.47$\pm$0.31& 0.04    &18.19$\pm$0.04&0.18&17.91$\pm$0.05&0.17     &2.3& ST\\
 2012/02/20.21    &5977.71&  +25.71& 19.25$\pm$0.07     & 0.04    &18.56$\pm$0.06& 0.03    &18.21$\pm$0.05&0.17&18.02$\pm$0.06&0.17     &2.5& ST\\
 2012/02/22.12    &5979.62&  +27.62& 19.29$\pm$0.03     & 0.05    &18.63$\pm$0.06& 0.02    &18.26$\pm$0.03&0.17&18.06$\pm$0.03&0.16     &2.1& ST\\
 2012/02/23.23    &5980.73&  +28.73&      $-$           & $-$     &18.56$\pm$0.07& 0.01    &18.35$\pm$0.09&0.17&18.18$\pm$0.04&0.15     &3.1& ST\\
 2012/03/03.12    &5989.62&  +37.62& 19.74$\pm$0.08     &$-0.05$  &18.76$\pm$0.03& 0.02    &18.33$\pm$0.03&0.20&18.30$\pm$0.25&0.16     &2.8& ST\\
 2012/03/11.05    &5997.55&  +45.55& 19.51$\pm$0.18     &$-0.16$  &18.74$\pm$0.19& 0.03    &18.23$\pm$0.07&0.23&18.18$\pm$0.04&0.17     &2.2& ST\\
 2012/03/14.09    &6000.59&  +48.59& 19.75$\pm$0.08     &$-0.12$  &18.74$\pm$0.05& 0.03    &18.16$\pm$0.08&0.23&18.07$\pm$0.03&0.17     &2.5& CT\\
 2012/03/23.23    &6009.73&  +57.73& $\textgreater19.55$&$-0.08$  &     $-$      & 0.04    &18.61$\pm$0.04&0.24&18.30$\pm$0.05&0.17     &2.4& ST\\
 2012/03/24.21    &6010.71&  +58.71& 19.74$\pm$0.07     &$-0.09$  &18.78$\pm$0.05& 0.04    &18.73$\pm$0.05&0.24&18.25$\pm$0.07&0.16     &2.0& ST\\
 2012/03/26.14    &6012.64&  +60.64& 20.06$\pm$0.05     &$-0.09$  &18.98$\pm$0.11& 0.05    &18.57$\pm$0.08&0.24&18.36$\pm$0.06&0.16     &2.8& ST\\
 2012/03/28.08    &6014.58&  +62.58& 21.04$\pm$0.11     &$-0.09$  &19.08$\pm$0.10& 0.05    &19.08$\pm$0.06&0.24&18.85$\pm$0.04&0.15     &1.5& CT\\
 2012/04/02.09    &6019.59&  +67.59& 21.54$\pm$0.15     &$-0.09$  &19.38$\pm$0.03& 0.05    &19.15$\pm$0.16&0.24&18.92$\pm$0.22&0.15     &3.1& ST\\
 2012/04/13.13    &6030.63&  +78.63&     $-$            & $-$     &     $-$      & $-$     &19.66$\pm$0.14&0.23&     $-$      & $-$     &3.1& ST\\
 2012/04/17.09    &6034.59&  +82.59& 21.85$\pm$0.09     &$-0.10$  &20.09$\pm$0.06& 0.06    &19.45$\pm$0.08&0.23&     $-$      & $-$     &2.6& CT\\
 2012/04/22.90    &6040.40&  +88.40&     $-$            &$-0.10$  &19.87$\pm$0.05& 0.06    &19.84$\pm$0.05&0.22&19.31$\pm$0.16&0.12     &1.9& NTT\\
 2012/04/26.02    &6043.52&  +91.52& 21.74$\pm$0.18     &$-0.11$  &20.01$\pm$0.12& 0.06    &19.87$\pm$0.06&0.22&19.57$\pm$0.08&0.11     &3.1& ST\\
 2012/04/28.01    &6045.51&  +93.51& 21.91$\pm$0.10     &$-0.12$  &19.90$\pm$0.06& 0.06    &20.16$\pm$0.13&0.22&19.55$\pm$0.10&0.10     &3.1& ST\\
 2012/04/29.97    &6047.47&  +95.47& 22.03$\pm$0.53     &$-0.12$  &20.20$\pm$0.10& 0.06    &20.22$\pm$0.08&0.22&19.57$\pm$0.12&0.09     &2.9& ST\\
 2012/04/30.89    &6048.39&  +96.39& 21.61$\pm$0.17     &$-0.12$  &20.31$\pm$0.08& 0.06    &19.90$\pm$0.09&0.22&     $-$      &     $-$ &1.7& NTT\\
 2012/05/13.98    &6061.48& +109.48& $\textgreater21.00$&$-0.13$  &20.13$\pm$0.09& 0.07    &20.33$\pm$0.20&0.20&19.65$\pm$0.13&0.07     &3.8& ST\\
 2012/05/24.88    &6072.38& +120.38& $\textgreater21.04$&$-0.14$  &20.53$\pm$0.12& 0.07    &20.44$\pm$0.38&0.20&19.96$\pm$0.21&0.05     &2.4& ST\\
\hline
\end{tabular}
\tablefoot{
  \tablefoottext{a}{Magnitude uncertainties are calculated after adding the
 $1\sigma$ uncertainty in photometry and the error associated with calibration in
 quadrature.}\newline
  \tablefoottext{b}{The event was dicovered on JD 2,455,956.04. However, the
 $V$-band maximum was estimated to be about 4\,d prior to discovery. Here the
 phases are calculated with respect to the epoch of $V$-band maximum in the
 observer frame, corresponding to JD 2,455,952. For details, see 
 Sect. \ref{phot}.}\newline
  \tablefoottext{c}{$K$-corrections are determined for all epochs with
 existing spectral information and by interpolating/extrapolating for other
 epochs.}\newline
  \tablefoottext{d}{FWHM of the stellar PSF in the $V$ band.
 }\newline
  \tablefoottext{e}{ST: 1.04\,m Sampurnanand Telescope + imaging Camera at ARIES, India; CT: 1.82\,m Copernico Telescope + AFOSC, Asiago Observatory; NTT: 3.6\,m
 New Technology Telescope + EFOSC2, ESO, Chile.}
}
\end{table*}
\begin{table*}
\setcounter{table}{4}
\centering
\caption{Journal of spectroscopic observations of SN 2012aa.\label{tab:speclog}}
\vskip 0.2cm
\scriptsize
\begin{tabular}{lc cc cc cc cc}
\hline\hline
{UT Date}&{JD $-$}&{Phase\tablefootmark{a}}&{Range}&{Telescope\tablefootmark{b}}&{Grating}&{Slit width}&{Dispersion}&{Exposure}&{S/N\tablefootmark{c}}\\
(yyyy/mm/dd)& 2,450,000 & (days)& (\mum)&     & (gr\,mm$^{-1}$)&  (\arcsec)& (\AA\,pix$^{-1}$)& (s)&(pix$^{-1}$)\\
\hline
        2012/02/02.14& 5959.64& +08 & 0.34$-$1.03& Lick    & 600/300  & 2.0 & 1.02/4.60& 1200&25\\
        2012/02/16.97& 5974.47& +22 & 0.36$-$0.90& IGO     & 300      & 1.0 & 1.4      & 2700&30\\
        2012/02/23.51& 5981.01& +29 & 0.34$-$1.00& Lick    & 600/300  & 2.0 & 1.02/4.60& 1350&25\\
        2012/03/12.84& 5999.34& +47 & 0.36$-$0.91& NTT     & 236      & 1.0 & 2.8      & 2700&35\\
        2012/03/15.14& 6001.64& +50 & 0.34$-$1.01& Keck-I  & 600/400  & 1.0 & 0.63/1.16&  600&30\\
        2012/04/25.80& 6043.30& +91 & 0.47$-$0.89& Gemini-S& 400      & 1.0 & 0.7      &  600&10\\
        2012/05/17.53& 6065.03&+113 & 0.34$-$1.01& Keck-I  & 600/400  & 1.0 & 0.63/1.16&  900&30\\
        2014/07/22.89& 6861.39&+909 & 0.37$-$0.79& NOT     & 300      & 1.0 & 3.0      & 3$\times$1500&28\\
\hline
\end{tabular}
\tablefoot{
  \tablefoottext{a}{With respect to the epoch of the $V$-band maximum in
 the observed frame, JD 2,455,952.}\newline
  \tablefoottext{b}{Lick: Kast on 3\,m Shane reflector; IGO: IFOSC on
  2\,m IUCAA Girawali Observatory, India; NTT: EFOSC2 on 3.6\,m New Technology
 Telescope, ESO, Chile; Keck-I: LRIS on 10\,m Keck-I telescope, Mauna Kea,
 Hawaii; Gemini-S: GMOS-S on 8.1\,m Gemini-South telescope, Chile; NOT:
 ALFOSC on 2.5\,m Nordic Optical Telescope, La Palma, Canarias, Spain.}\newline
  \tablefoottext{c}{Signal-to-noise ratio at 0.6\,\mum.}
}
\end{table*}

 We have also used the unfiltered data taken from CRTS, which provide extensive
 coverage of the field. All the unfiltered CRTS magnitudes have been calibrated
 to the Johnson $V$ band using the local sequences. The CRTS images have low
 spatial resolution (2\farcs43 pixel$^{-1}$). Photometry of all CRTS
 observations was performed in two different ways: {\it (i) Flux subtraction:} A
 calibrated template frame was produced after adding 10 aligned CRTS images
 having moderate seeing ($\sim$2\farcs5) and without the SN. Then the calibrated
 flux of the host from this ``stacked template'' was subtracted from all other
 CRTS measurements, in the same way as mentioned above for the photometry of
 other observations. {\it (ii) Template subtraction:} This was performed by
 using the ``stacked template'' as a reference image, following standard
 procedures (e.g., \citealt{2011MNRAS.414..167R}, and references therein).
 Comparison of the results of these two processes shows consistency (see Sect.
 \ref{phot}). Photometry with flux subtraction is presented in Table
 \ref{tab:crtslog}\footnote{Table \ref{tab:crtslog} is available only in the
 Online Journal.}.

\subsection{Low-resolution optical spectroscopy} \label{obs:spec}
 Long-slit low-resolution spectra ($\sim 6$--14\,\AA) in the optical range
 (0.35--0.95\,\mum) were collected at eight epochs between +12\,d and +905\,d
 after discovery. These include two epochs from the 3\,m Shane reflector at
 Lick Observatory, two epochs from the 10\,m Keck-I telescope, and one epoch
 each from the 2\,m IUCAA Girawali Observatory (IGO, India), the 2.56\,m Nordic
 Optical Telescope (NOT, La Palma), the 3.6\,m NTT, and the 8\,m Gemini-South
 telescope. All of the spectroscopic data were reduced under the IRAF
 environment\footnote{Bias and flat-fielding were performed on all frames.
 Cosmic rays were rejected with the Laplacial kernel detection method
 \citep{2001PASP..113.1420V}. Flux calibration was done using spectrophotometric
 fluxes from \citet{1994PASP..106..566H}. For a few cases where we did not
 observe any spectrophotometric standard stars, relative flux calibrations were
 performed.}. All spectra of \sn\ have been normalized with respect to the
 average continuum flux derived from the line-free region 6700--7000\,\AA. For
 the rest of the analysis we have used the normalized spectra of the SN. The
 spectrum, obtained with NOT+ALFOSC on 2014 July 22 (+905d after discovery),
 contains mainly emission from the host. This flux-calibrated spectrum was
 utilized to quantify the host properties. The journal of spectroscopic
 observations is given in Table \ref{tab:speclog}.

\section{Distance and extinction toward \sn} \label{DistExt} 
 The distance was calculated from the redshift of the host galaxy. The host
 spectrum (see Sect. \ref{Host}, Fig. \ref{fig:Hostspec}) contains strong
 emission lines of \ha, \hb, \Niia\,$\lambda\lambda$6548, 6583,
 \Oiia\,$\lambda$3727, and \Siia\,$\lambda\lambda$6717, 6731. Using these
 emission lines, the measured value of the redshift is $0.0830\pm0.0005$. This
 corresponds to a luminosity distance $D_L =380\pm2$\,Mpc using a standard
 cosmological model\footnote{The cosmological model with $H_0$ = 69.6
 \kms\,Mpc$^{-1}$, $\Omega_{m}$ = 0.286, and $\Omega_{\Lambda}$ = 0.714
 \citep{2014ApJ...794..135B} is assumed throughout the paper.}, and hence the
 distance modulus ($\mu$) is $37.9\pm0.01$\,mag.

 The extinction toward \sn\ is dominated by the Milky Way contribution. The
 Galactic reddening in the line of sight of \sn, as derived by
 \citet{2011ApJ...737..103S}, is \ebv\ = $0.088\pm0.002$ mag. The host
 contribution toward the SN can be estimated from the \Nai\,D absorption line
 in spectra of the SN. We use the empirical relation between the equivalent
 width of \Nai\,D absorption and \ebv\ (\citealt{2012MNRAS.426.1465P}, and
 references therein).

 In the SN spectra we did not detect any significant \Nai\,D absorption at the
 rest wavelength of the host. The dip nearest to this wavelength has an
 equivalent width of 0.30\,\AA\ and comparable to the noise level. This would
 correspond to \ebv\ $\approx0.032$\,mag and can be considered an upper limit 
 for the contribution of the host to the reddening toward the SN.

 Hence, an upper limit for the total reddening toward \sn\ is \ebv$_{\rm upper}
 = 0.12$\,mag, while its lower limit is \ebv$_{\rm lower} = 0.088$\,mag.
 We use \ebv$_{\rm upper}$ as the total reddening toward the SN. This
 corresponds to a total visual extinction $A_V = 0.37$\,mag, adopting the 
 ratio of total-to-selective  extinction $R_V = 3.1$.

\section{Light-curve evolution of \sn} \label{phot} 
\begin{figure}
\centering
\includegraphics[width=8.5cm]{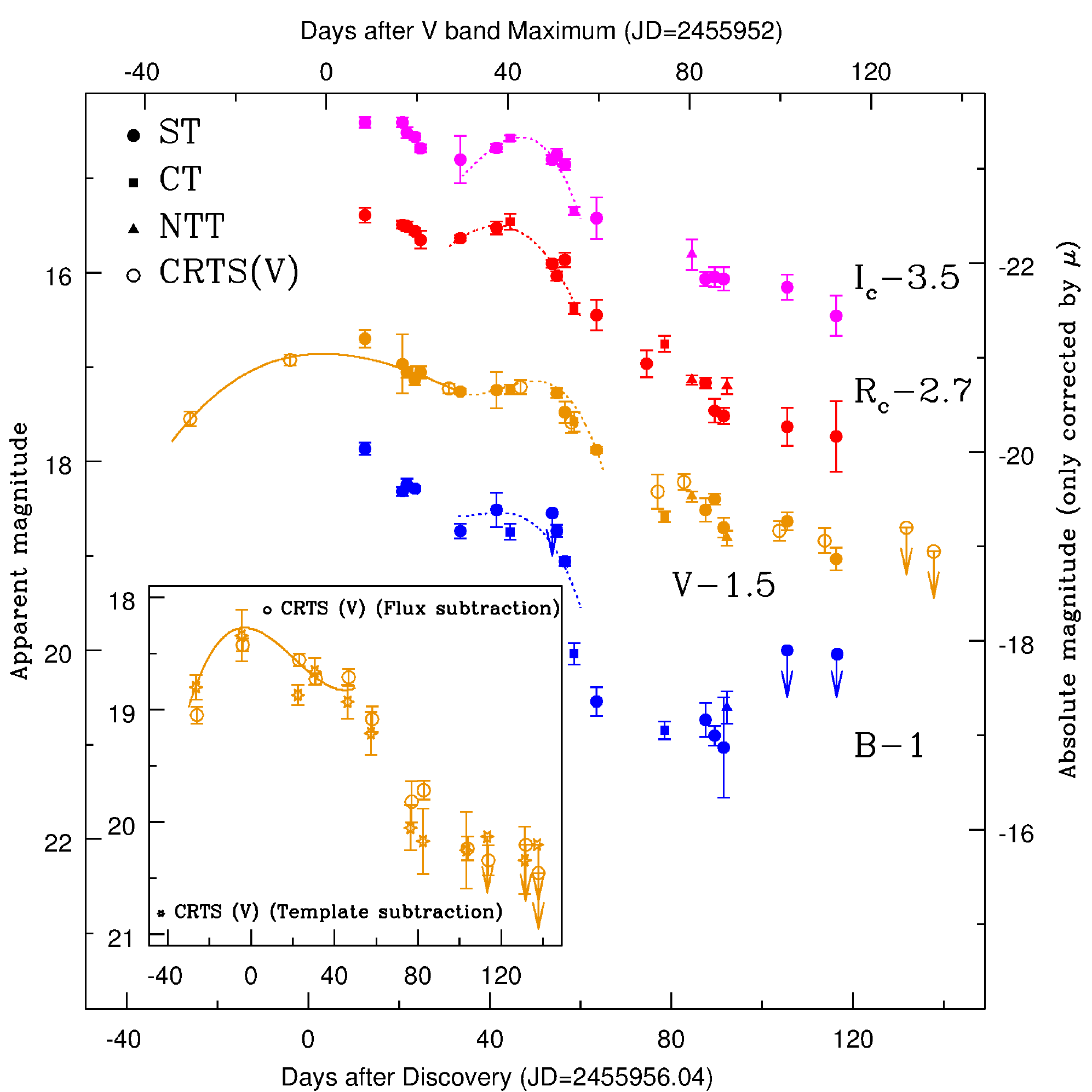}%
\caption{The optical light curves of SN 2012aa. The $BVR_cI_c$ photometric
 measurements from ST, CT, and NTT are shown respectively with filled
 circles, squares, and triangles, while the CRTS measurements are open circles.
 The epochs are given with respect to the date of discovery. All the unfiltered
 CRTS magnitudes have been calibrated to the Johnson $V$-filter system using 
 the local sequences. The solid line represents a third-order polynomial fit 
 around the first broad peak observed in $V$, while the dotted lines represent
 third-order polynomial fits around the secondary peak in $BVR_cI_c$. The
 inset represents a comparison between the measurements from CRTS observations
 by using ``template subtraction'' (open stars) and ``flux subtraction''
 (open circles). The solid line around the peak is again a third-order
 polynomial fit around the first peak using the measurements obtained 
 from template subtraction. It is worth noting that the absolute-magnitude 
 scale is corrected only for distance modulus ($\mu$) to show the approximate 
 luminosity. When determining a more accurate luminosity, reddening and 
 the $K$-correction were also considered. See text for details.}
\label{fig:applc}
\end{figure}

 The optical photometric light curves are shown in Fig. \ref{fig:applc} and
 given in Table \ref{tab:photlog}.

 The long-term photometry of the host recorded by CRTS is presented in Fig.
 \ref{fig:crtslc}. The photometric measurements of the CRTS images by the
 flux-subtraction and template-subtraction methods give consistent results (see
 the inset of Fig. \ref{fig:applc}). To maintain consistency in the reductions
 of all photometric observations, we adopt the results of the flux-subtraction
  method in the rest of the work.

 From the CRTS data it is clear that the explosion took place long before the
 discovery. \sn\ was actually discovered around its peak brightness. However,
 owing to the very sparse dataset before the peak, precise estimates of the
 epoch of maximum and the corresponding peak magnitude using polynomial fitting
 are difficult. A third-order polynomial fit to the $V$-band peak magnitudes
 (solid-line in Fig. \ref{fig:applc}) indicates that the epoch of $V$-band
 maximum was JD 2,455,952\footnote{For the rest of the work, all of the phases
 (if not stated otherwise) have been estimated with respect to JD 2,455,952
 measured in the observer's frame.}, with $M_V \approx -19.87$\,mag. The host
 $V$-band magnitude, measured from pre-SN CRTS data, is 19.11 mag (see Sect.
 \ref{phot_long}). The polynomial fits to the combined $V$-band and CRTS data of
 \sn\ reach 19.11\,mag roughly between 30 -- 35\,d prior to the discovery.
 Therefore, hereafter we assume that the event had a long rise time of $\gtrsim
 30$\,d.

\subsection{Early evolution ($\textless 60$ days after peak)} \label{phot_early}
 A slowly rising, broad light curve with a secondary peak makes this event
 special. The $V$-band light curve seems to rise to maximum on a similar time
 scale as it later declines after peak. Moreover, by fitting third-order
 polynomials to the \bvri\ light curves after +30\,d, we find a rebrightening
 in all optical bands that produce a second peak (or bump) in the light curves,
 appearing between $\sim$ +40\,d and +60\,d after the first $V$-band maximum.
 Beyond that, a nearly exponential decline of the optical light marks the onset
 of the light-curve tail. This kind of secondary bump in the light curve has
 been observed in a few other CCSNe, but is not common in SNe~Ic. Different 
 physical processes could explain these data, and a few scenarios are discussed
 below.

 Objects that resemble \sn\ photometrically are the Type Ibn/IIn SNe 2005la
 \citep{2008MNRAS.389..131P} and 2011hw \citep{2012MNRAS.426.1905S,
 2015MNRAS.449.1921P} and the SN~Ibn iPTF13beo \citep{2014MNRAS.443..671G},
 where double or multiple-peaked light curves were also reported. To
 some extent the light curves of the luminous Type IIn (or Type IIn-P) SNe 1994W
 \citep{1998ApJ...493..933S} and 2009kn \citep{2012MNRAS.424..855K} also
 exhibited similar features, with relatively flat light curves during their
 early evolution. All of these were interpreted as being produced by 
 SN shock interaction with the dense CSM. This proposition was supported by the
 appearance of strong, narrow H/He features in the SN spectra, caused by
 electron scattering in the dense CSM during the interaction process
 (\citealt{2001MNRAS.326.1448C, 2012MNRAS.426.1905S, 2014ApJ...797..118F}, and
 references therein). The pre-explosion outburst in the progenitor is the
 potential mechanism that can produce such dense CSM
 \citep{2014ARA&A..52..487S}. In fact, in a few luminous blue variables (LBVs)
 such pre-explosion outbursts have been recorded (e.g.,
 \citealt{1999PASP..111.1124H, 2010AJ....139.1451S, 2013ApJ...767....1P,
 2013ApJ...779L...8F, 2014ApJ...789..104O}, and references therein). During its
 major event in 2012, the progenitor of SN 2009ip also exhibited a bump
 $\sim40$\,d after its first maximum which was interpreted as an effect of shock
 interaction (\citealt{2014MNRAS.438.1191S, 2014ApJ...787..163G}; see also
 \citealt{2015AJ....149....9M}, and references therein). Similarly, SN 2010mb,
 which is spectroscopically very similar to \sn, is a rare example where
 interaction of a SN~Ic with an H-free dense CSM was proposed
 \citep{2014ApJ...785...37B}. 

 However, the spectral sequence of \sn\ (see Sect. \ref{spec}) shows that
 the effects of shock interaction are not prominent in its line evolution.
 The photometric behaviour of the SN must therefore be caused mainly by 
 changes in the continuum flux. 

 Recently, \citet{2015MNRAS.452.3869N} made a comparative study between
 SLSNe~I and SNe~Ic with a sample of 21 objects. They computed the rise
 ($\tau_{\rm ris}$) and decline ($\tau_{\rm dec}$) timescales, which were
 defined as the time elapsed before (or after) the epoch of maximum during which
 the SN luminosity rises to (or declines from) the peak luminosity from (or to) 
 a value that is $e^{-1}$ times its peak luminosity. This corresponds to a drop
 of 1.09\,mag. Using these definitions and by fitting third-order polynomials
 around the first $V$ maximum, we found $\tau_{\rm ris} \approx 31$\,d and
 $\tau_{\rm dec} \approx 55$\,d for \sn. However, unlike the SLSNe discussed by
 \citet{2015MNRAS.452.3869N}, \sn\ exhibits light curves with shallow decline
 after the peak followed by a secondary bump before entering the nebular phase.
 Therefore, we calculated the post-maximum decline rate ($R_{\rm dec}$
 [mag\,d$^{-1}$]) of this object in the rest frame to compare the nature of its
 light curves with those of other SNe. Here we simply assume that the secondary
 bump at $\sim$ +50\,d is produced by another physical process.

 For events (e.g., H-poor CCSNe) with $T_{\rm neb} \gtrsim \tau_{\rm dec}$, we
 can consider $\tau_{\rm dec} \approx R_{\rm dec}^{-1}$, where $T_{\rm neb}$ is
 the time (with respect to the time of maximum brightness) when the tail phase
 starts. The post-maximum decline rates ($R_{\rm dec}$) of \sn\ in the \bvri\
 bands (calculated between +15\,d and +30\,d) are $\sim0.03$, 0.02, 0.01, and
 0.02\,mag\,d$^{-1}$, respectively. These values are less than $R_{\rm fall}$
 and comparable to $R_{\rm tail}$ of broad-lined SNe~Ic\footnote{$R_{\rm fall}$
 and $R_{\rm tail}$ are the values of $R_{\rm dec}$ calculated before and after
 $T_{\rm neb}$. The values of $R_{\rm fall}$ in the \bvri\ bands for broad-lined
 SNe~Ic such as SNe 1998bw ($\sim 0.11$, 0.07, 0.06, and 0.04\,mag\,d$^{-1}$),
 2002ap ($\sim 0.09$, 0.06, 0.05, and 0.04\,mag\,d$^{-1}$), 2003jd ($\sim 0.09$,
 0.08, 0.07, 0.06\,mag\,d$^{-1}$), and 2006aj ($\sim 0.12$, 0.07, 0.06,
 0.05\,mag\,d$^{-1}$) are higher than $R_{\rm tail}$
 ($\sim 0.02$\,mag\,d$^{-1}$ in the \bvri\ bands for all above-mentioned
 CCSNe).}. The post-maximum decline rates of \sn\ in the different bands are
 consistent with the values of $\tau_{\rm dec}$ obtained by
 \citet{2015MNRAS.452.3869N} using the quasi-bolometric light curves of CCSNe,
 and are also similar to the SLSNe(-I) which decline fast after maximum
 brightness (e.g., SNe 2010gx, 2011ke, 2011kf, PTF10hgi, PTF11rks), but not
 comparable to SLSNe(-R) having a slow decline after maximum (e.g., SNe 2007bi,
 2011bm, PTF12dam, PS1-11ap). We will discuss this further in Sect.
 \ref{res:comp}.

 \subsubsection{The secondary bump in \sn}\label{sec:bump}
 A second bump in the SN light curve beyond maximum brightness is not common 
 among SNe~Ibc. In \sn\ it was observed at $\sim 50$\,d after the principal peak
 (i.e., $\sim80$\,d after the explosion, assuming a rise time of
 $\gtrsim30$\,d). In the simplest scenario, this secondary bump may be an effect
 of shock interaction with clumpy CSM having a jump in its density profile.
 Under the assumption that the typical SN shock velocity is $\sim10^4$\,\kms,
 the region of the density jump should be located at a distance of
 $\sim10^{16}$\,cm from the explosion site.

 If the density jump is spherically symmetric around the SN location, it may be
 in the form of a dense shell, created from an eruption by the progenitor before
 the SN explosion. If we consider that the outward radial velocity of the
 erupted material is $\sim10^3$\,\kms, we can estimate that in \sn, the putative
 pre-SN stellar eruption happened a few years prior the final explosion. In the
 existing pre-SN data on \sn\ (Sect. \ref{phot_long}) we could not find any 
 significant rise in luminosity due to outburst as a precursor activity (for
 e.g., see \citealt{2010AJ....139.1451S}). This is, however, not surprising
 given that our observations are insensitive to the expected small variations
 and the stellar eruptions are of short duration, quite unpredictable, and may
 happen at any time.

 A secondary bump is also observed in the near-infrared light curves of Type Ia
 SNe, to some extent similar to the secondary bump in \sn. In SNe~Ia this is
 caused by the transition of \Feiii\ ions to \Feii\ ions and observable mainly
 at longer wavelengths \citep{2006ApJ...649..939K}. Unlike SNe~Ia, \sn\ exhibits
 a secondary bump also in the $B$ and $V$ bands. This indicates that the
 generation mechanism for the secondary bumps in \sn\ is different from that in
 SNe~Ia.
\begin{figure}
\centering
\includegraphics[width=8.5cm]{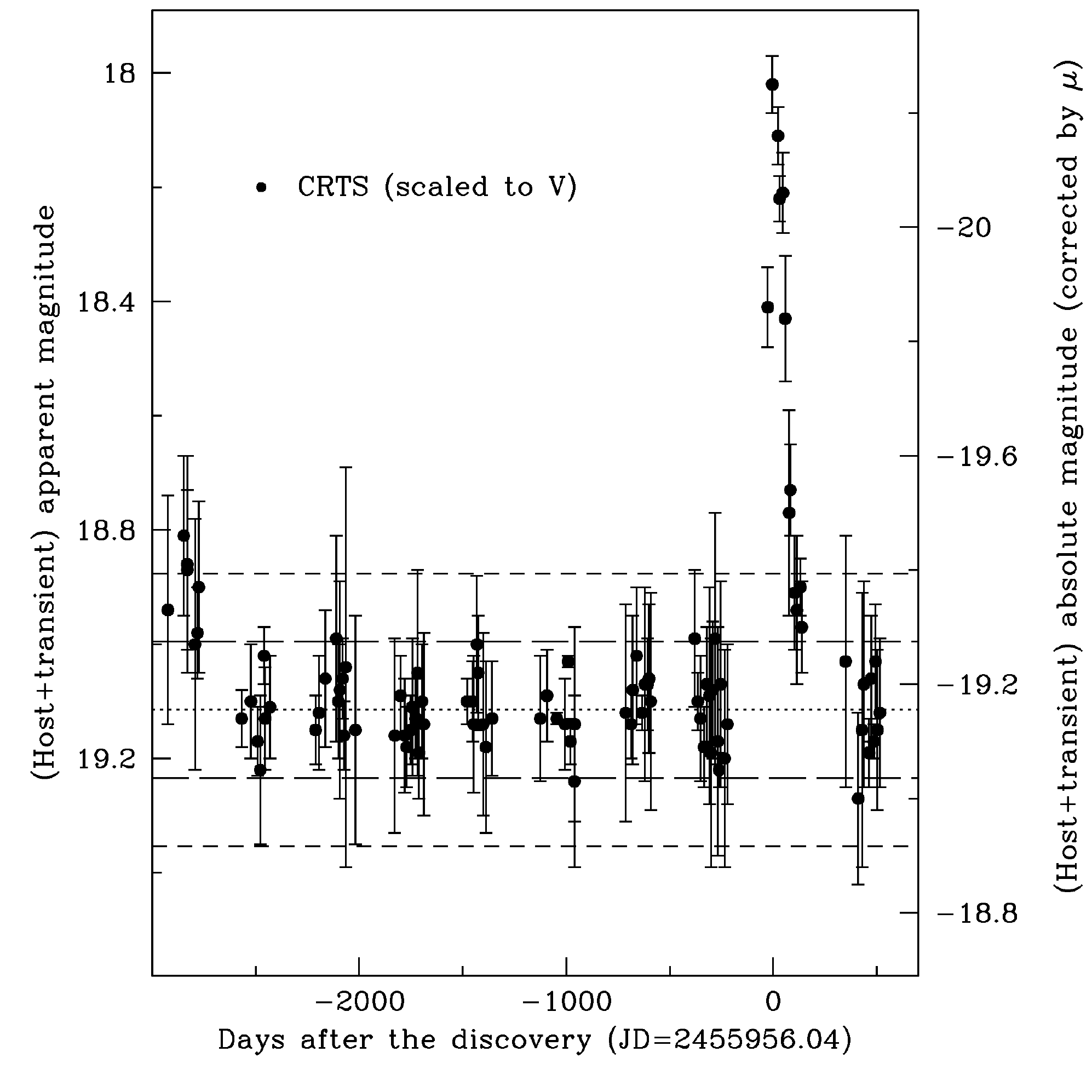}%
\caption{Long-term  monitoring of the host of \sn. The CRTS data points
 represent photometric observations of the system (host + transient)
 from $\sim8$\,yr prior the burst to $\sim1.5$\,yr after the burst. The
 dotted line represents the weighted (and scaled to $V$) mean magnitude
 ($19.11\pm0.24$) of the host, calculated after excluding measurements
 when the transient was visible. Long-dashed lines are 1$\sigma$ limits
 and short-dashed lines are 2$\sigma$ limits.}
\label{fig:crtslc}
\end{figure}

\subsection{Late evolution ($\textgreater 60$ days post peak)} \label{phot_late}
 The SN flux started to decrease after $\sim+55$\,d, showing roughly an
 exponential decay beyond +70\,d in all bands. We observed a relatively larger
 drop ($\sim 2$\,mag) in the $B$-band flux at the end of the plateau in
 comparison to the $VR_cI_c$ bands ($\sim 1$\,mag). This is consistent with the
 decrement of the blue continuum in late-time ($\textgreater+50$\,d) spectra.
 We calculated the decline rate at this phase in the rest frame in all bands.
 Although the observations are sparse, linear fits to the light curves at
 $\textgreater+70$\,d give the following decline rates [in mag\,d$^{-1}$]:
 $\gamma_{V} \approx 0.014$, $\gamma_{R} \approx 0.025$, and $\gamma_{I}
 \approx 0.014$. Certainly this implies that the bolometric flux also declines
 rapidly (see Sect. \ref{ColBol}). The observed rates are faster than that of
 $^{56}{\rm Co}\rightarrow ^{56}$Fe (0.0098 mag\,d$^{-1}$), though consistent
 with the post-maximum decline rates of SLSN-R events and comparable to the
 decline rates ($R_{\rm tail}$) of SNe~Ibc.
\begin{figure*}
\centering
\includegraphics[width=8.5cm]{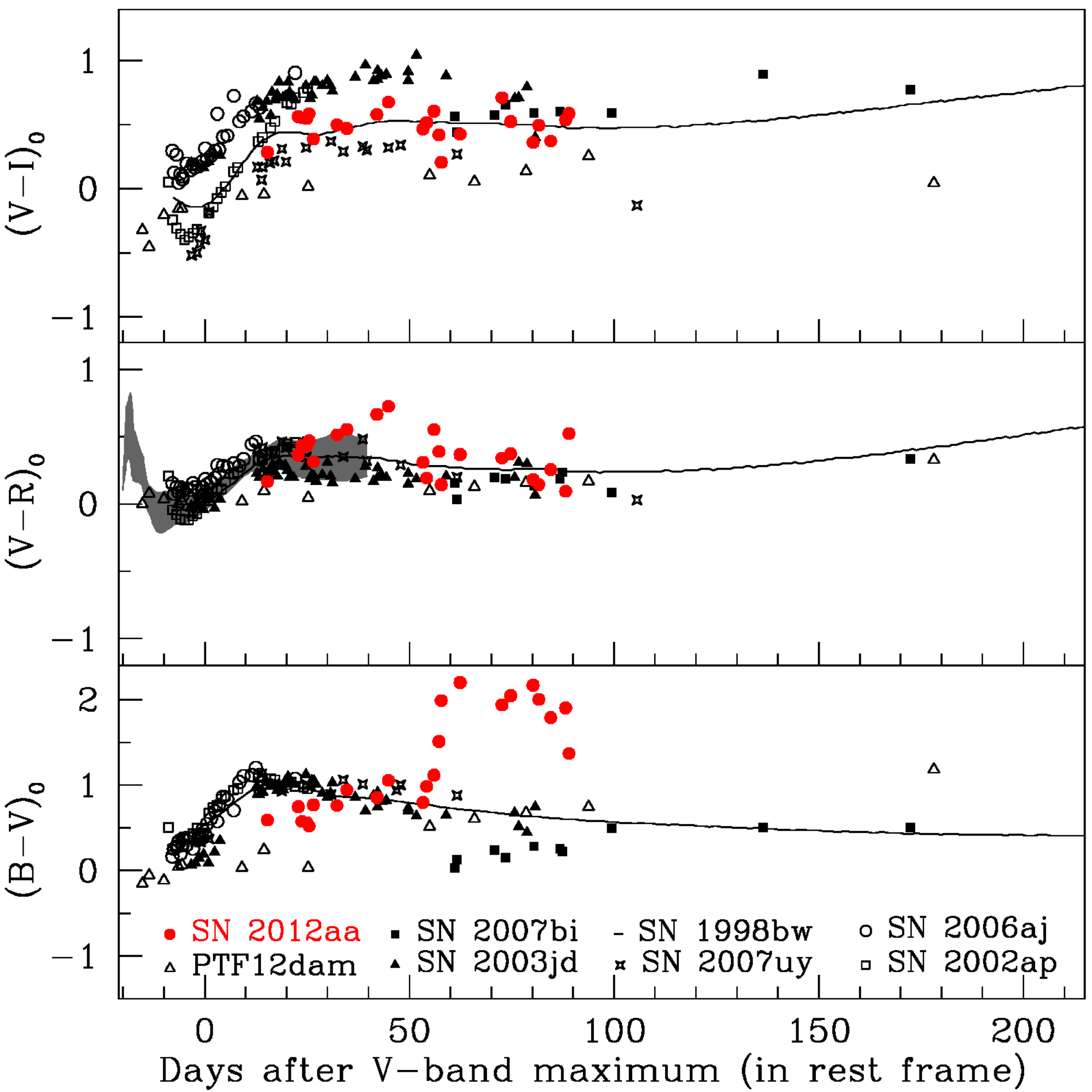}
\hskip 0.5cm
\includegraphics[width=8.5cm]{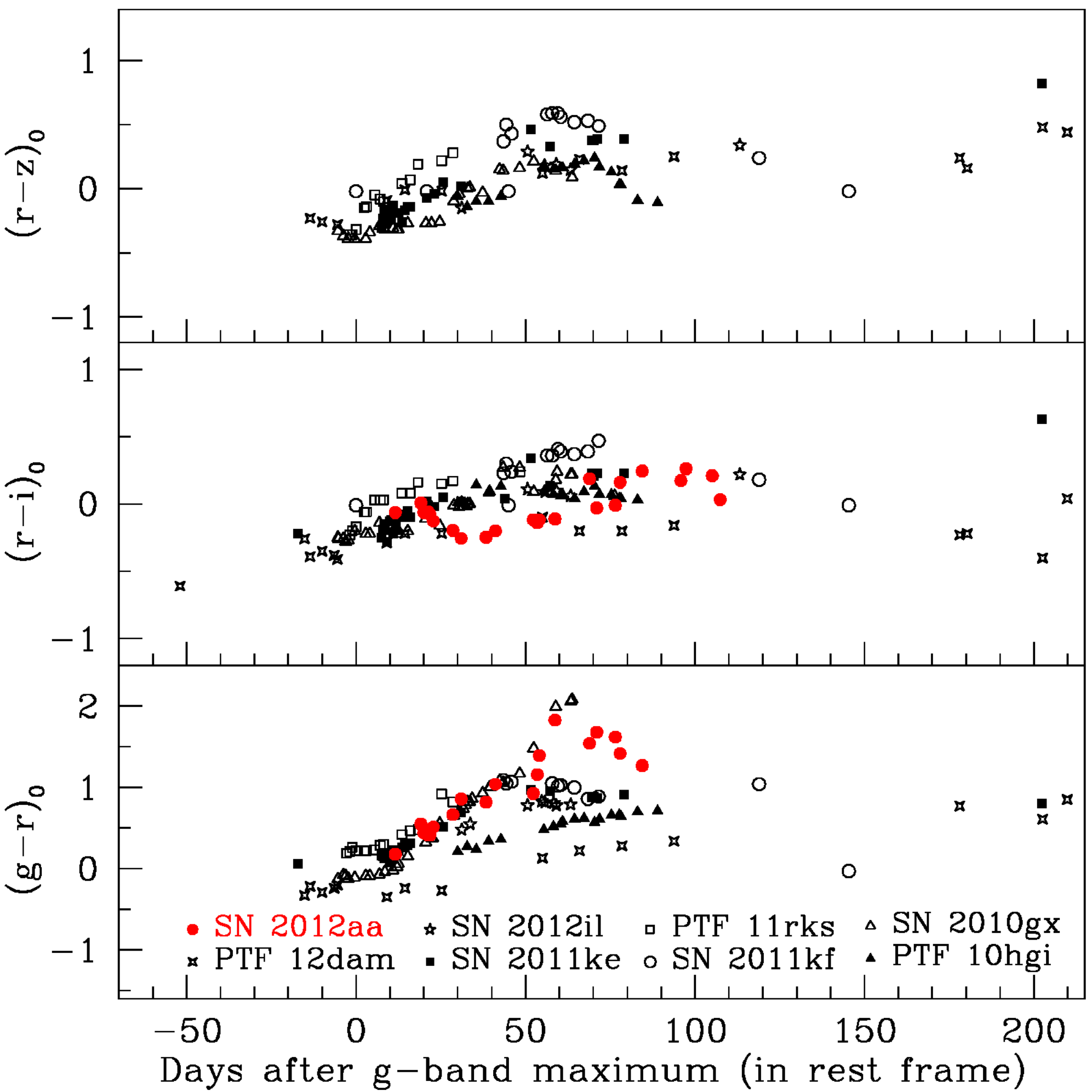}
\caption{ Colour evolution of \sn\ and comparison with other events. All of 
 the colours have been corrected for reddening. {\bf Left panel:}
  four broad-lined SNe~Ic [SNe 1998bw
 \citep{2011AJ....141..163C}, 2002ap \citep{2003PASP..115.1220F}, 2003jd
 \citep{2008MNRAS.383.1485V}, and 2006aj \citep{2006A&A...457..857F}],  
 along with the normal SN~Ibc SN 2007uy \citep{2013MNRAS.434.2032R} and 
 two SLSNe~Ic [SN
 2007bi \citep{2010A&A...512A..70Y} and PTF12dam \citep{2013Natur.502..346N}]
 are compared with the colour evolution of \sn. The shaded region in the
 central panel ($(V-R)_0$ colour) represents the colour evolution
 of SNe~Ibc as discussed by \citet{2011ApJ...741...97D}. Unlike normal
 and broad-lined SNe~Ibc, \sn\ and SLSNe show evolution with longer
 timescales. {\bf Right panel:} the comparison with SLSNe [SNe 2012il,
 2011ke, 2011kf, PTF 11hgi, and 11rks from \citep{2013ApJ...770..128I}; 
 PTF 12dam from \citep{2013Natur.502..346N}; and SN 2010gx from
 \citep{2010ApJ...724L..16P}]. After maximum brightness, the colour of
 SNe~Ibc appears to evolve quickly, whereas in SLSNe it evolves more slowly,
 becoming redder over a longer time. The sudden increment in the $(B-V)_0$
 (or $(g-r)_0$) colour of \sn\ after +50\,d is similar to the colour variation
 observed in SN 2010gx.
}
\label{fig:colour}
\end{figure*}

\subsection{Long-term photometry of the host galaxy} \label{phot_long}
 Figure \ref{fig:crtslc} presents the long photometric monitoring of the host
 galaxy by CRTS, starting from $\sim2923$\,d before the discovery of \sn\ to
 $+515$\,d after discovery. The weighted mean magnitude of the host in $V$,
 indicated by the dotted line, is $19.11\pm0.24$; the long-dashed lines are
 1$\sigma$ and the short-dashed lines are 2$\sigma$ limits. This value is
 consistent with the $V$-band measurement obtained years after the explosion
 (Sect. \ref{obs:phot}). The peak brightnesses of typical precursor events in
 massive stars varies roughly between $-10$ and $-15$\,mag
 \citep{2010AJ....139.1451S, 2010MNRAS.408..181P, 2011MNRAS.415..773S}.
 At the distance of \sn, this corresponds to an apparent brightness of
 $\sim22$--25\,mag --- undetectable in the CRTS long-term monitoring. 

 In some of the very early-time ($\sim2900$\,d) observations prior to the SN, we
 found that the measurements are marginally higher than the 2$\sigma$ limit.
 Subtracting the mean flux of the host, this would correspond to a rise to
 $\sim20.75$\,mag, or $\sim-17.59$\,mag on the absolute scale. This is about
 3--7\,mag more luminous than typical pre-SN outbursts of LBVs, and comparable
 only to the precursor activity of SN 1961V. If real, this early event caught by
 CRTS is more likely associated with some other SN that happened in the same
 host or to nuclear activity in the host itself. However, note that these early
 CRTS data were taken when the CSS-CCD system response was different\footnote{
 The CSS-CCD camera used during the observations taken between JD 2,453,032 and
 JD 2,453,182, roughly between 2924\,d and 2775\,d prior to the SN discovery,
 was not thinned. A thinned CCD has been used for observations of the \sn\ field
 after JD 2,453,182.}. This may also be a cause for getting a higher flux for
 the object at early epochs. 

\section{Color Evolution and Quasi-Bolometric Flux} \label{ColBol} 
 Figure \ref{fig:colour} presents the intrinsic colour evolution of \sn\ and a
 comparison with other SNe~Ic and SLSNe. The left panel draws the comparison
 with SNe~Ibc (e.g., SNe 1998bw, 2002ap, 2003jd, 2006aj, 2007uy) and Type I
 SLSNe (e.g., 2007bi, PTF12dam) in the Bessell system, while for better
 statistics a separate comparison between \sn\ and Type I SLSNe is shown in the
 right panel using Sloan colours. The Sloan colours of \sn\ have been calculated
 after converting Bessell magnitudes to Sloan magnitudes through
 ``$S$-corrections'' using the filter responses given by
 \citet{2014A&A...562A..17E}. 

 The difference in colours between \sn\ and other similar events (e.g., SN
 2003jd, PTF12dam) at early epochs is below 0.5\,mag. If we assume they have
 similar intrinsic colours, this confirms that our estimate of the
 line-of-sight extinction toward \sn\ is reasonable and thus establish that \sn\
 is indeed an intermediate-luminosity transient between CCSNe and SLSNe in terms
 of its peak absolute magnitude.

 With a larger sample of SNe~Ibc, \citet{2011ApJ...741...97D} (also see the left
 panel of Fig. \ref{fig:colour}) found that the \cvr\ colours of these SNe
 become redder (an effect of cooling; \citealt{2013ApJ...769...67P,
 2015A&A...574A..60T}) within 5\,d after explosion. Then they rapidly become
 bluer just before or around the peak ($\sim20$\,d after explosion), and
 subsequently redder again starting 20\,d post maximum. In contrast, the
 evolutionary timescale of SLSNe~I is much longer. The \cgr, \cri, and \crz\
 colours evolve gradually from negative/zero values at around peak brightness to
 positive values (+0.5--+1\,mag) by +50\,d. Owing to a lack of data, both the
 early-time (before maximum brightness) and late-time (beyond +50\,d) colour
 evolution of SLSNe is not well known. The few existing early-time and
 late-time observations of the Type I SLSNe 2011ke, 2011kf, and PTF12dam
 indicate that these SNe get bluer gradually well before their maxima, and keep
 roughly constant colour values beyond +50\,d.

 Since we do not have any multiband observations of \sn\ before peak, from the
 left panel of Fig. \ref{fig:colour} it is difficult to assess whether this SN
 behaves like other CCSNe or not. Between +10\,d and +50\,d, the \cbv\ colours
 of \sn\ are more similar to those of CCSNe rather than those of the Type I
 SLSNe 2007bi and PTF12dam. The difference is negligible (and hence not
 conclusive) in \cvr\ and \cvi\ colours. A further comparison of \sn\ with a
 larger sample of SLSNe is presented in the right panel of Fig.
 \ref{fig:colour}. The increase in \cgr\ of \sn\ is similar to that of SNe
 2010gx, 2011ke, 2011kf, and PTF11rks. Like SN 2010gx, \sn\ also shows a jump in
 the \cgr\ (and in \cbv) colour after +50\,d, owing to a steeper decline in the
 blue filter.
\begin{figure}
\includegraphics[width=8.5cm]{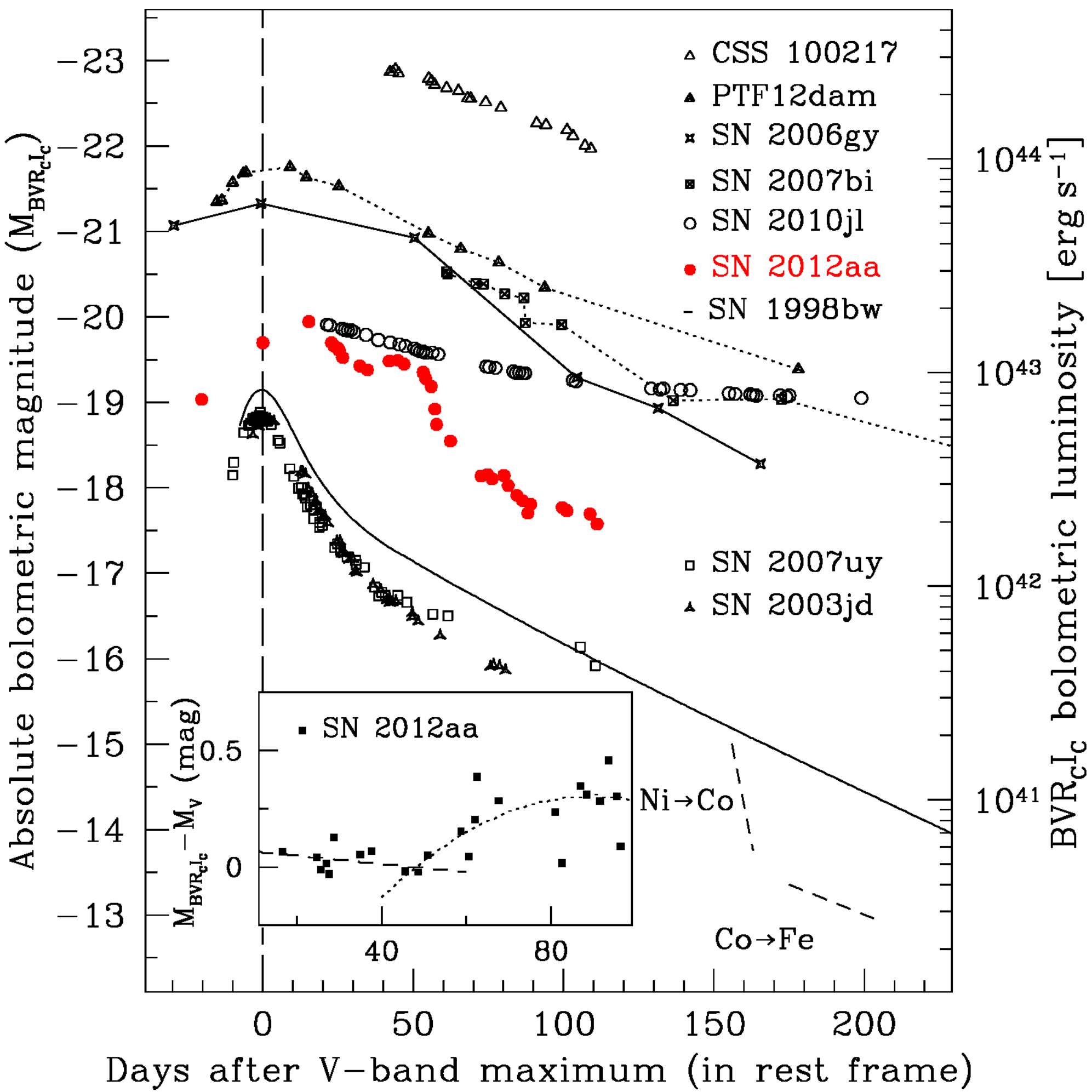}%
\caption{The \bvri\ quasi-bolometric light curve of SN 2012aa and other
 stripped-envelope CCSNe and SLSNe along with the luminous event SN 2010jl.
 The long dashed vertical line shows the epoch of $V$-band maximum light. The
 $K$-correction and cosmological time dilation effects
 have been taken into account. For the first two data points, we have only CRTS
 $V$-band data, while for the last four points there is incomplete
 coverage/detection in some bands. For these six epochs we have calculated
 the $BVR_cI_c$ quasi-bolometric light curves after measuring the temporal
 evolution of ($M_{BVR_cI_c} - M_V$) as shown in the inset.}
\label{fig:bol} 
\end{figure}

 The \bvri\ quasi-bolometric light curve has been constructed after accounting
 for the line-of-sight extinction, K-correction, and cosmological time dilation.
 A comparison of \sn\ with other SLSNe and CCSNe is presented in Fig.
 \ref{fig:bol}.\footnote{Quasi-bolometric luminosities of \sn\ have been
 calculated for those epochs where the SN was detected in at least in two bands.
 For the epochs when SN was not observed in some band, the absolute magnitudes
 in corresponding bands were calculated by linear interpolation. It should be
 noted that for the first two epochs, we have only CRTS data, while for the last
 four points there is incomplete coverage/detections in all \bvri\ bands. For
 these 6 epochs we calculated the \bvri\ quasi-bolometric luminosities after
 measuring the temporal evolution of ($M_{BVR_cI_c}$--$M_V$) as shown in the
 inset of Fig. \ref{fig:bol}. Two different polynomials have been fitted to the
 dataset, calculated on the epochs for which $M_{BVR_cI_c}$ and $M_V$ were
 known. After excluding four deviant points we get two following fits: for
 $t\textless50$\,d, $(M_{BVR_cI_c}-M_V) = 0.0862 - 0.0018\,t$, while for
 $t\textgreater50$\,d, $(M_{BVR_cI_c}-M_V) = -1.1125 + 0.0315\,t
 - 0.00016\,t^2$, where $t$ is the time elapsed after $V$-band maximum. The
 first fit was used to calculate $M_{BVR_cI_c}$ for the first two data points,
 while the second fit was used to get $M_{BVR_cI_c}$ for the last four epochs.}.
 Fitting a third-order polynomial to the early part of the quasi-bolometric
 light curve  sets the epoch of bolometric maximum about 7\,d after $V$-band
 maximum, as mentioned in Sect. \ref{phot}, or just 4\,d after discovery. The
 rise and decline timescales for the quasi-bolometric light curve are
 respectively 34\,d and 57\,d, consistent with the corresponding values
 calculated from the $V$ band (see Sect. \ref{phot_early}).

 In Fig. \ref{fig:bol} we compare the \bvri\ quasi-bolometric light curve of
 \sn\ with those of SNe~Ibc (e.g., SNe 1998bw, 2003jd, 2007uy) and the SLSN 
 PTF12dam. For completeness, we also include the H-rich SLSNe 2006gy
 \citep{2009ApJ...691.1348A} and CSS100217 \citep{2011ApJ...735..106D}, as well
 as the long-lived luminous transient Type IIn SN 2010jl
 \citep{2012AJ....144..131Z}. It is clear that in terms of both peak luminosity
 and light-curve width, \sn\ is intermediate between SNe~Ic and SLSNe I. It
 probably belongs to the class of luminous transients that are supported to
 some extent by an additional powering mechanism along with radioactive decay.
 In particular, the bumpy light curve strongly supports the CSM-interaction
 scenario. Between +60\,d and +100\,d after maximum brightness, the decline rate
 of the quasi-bolometric light of \sn\ is 0.02 mag\,d$^{-1}$, steeper than the
 $^{56}{\rm Co}\rightarrow^{56}$Fe rate though consistent with the decline rates
 of SNe~Ibc. Beyond +100\,d the light curve of \sn\ became flatter.

\begin{figure}
\centering
\includegraphics[width=8.5cm]{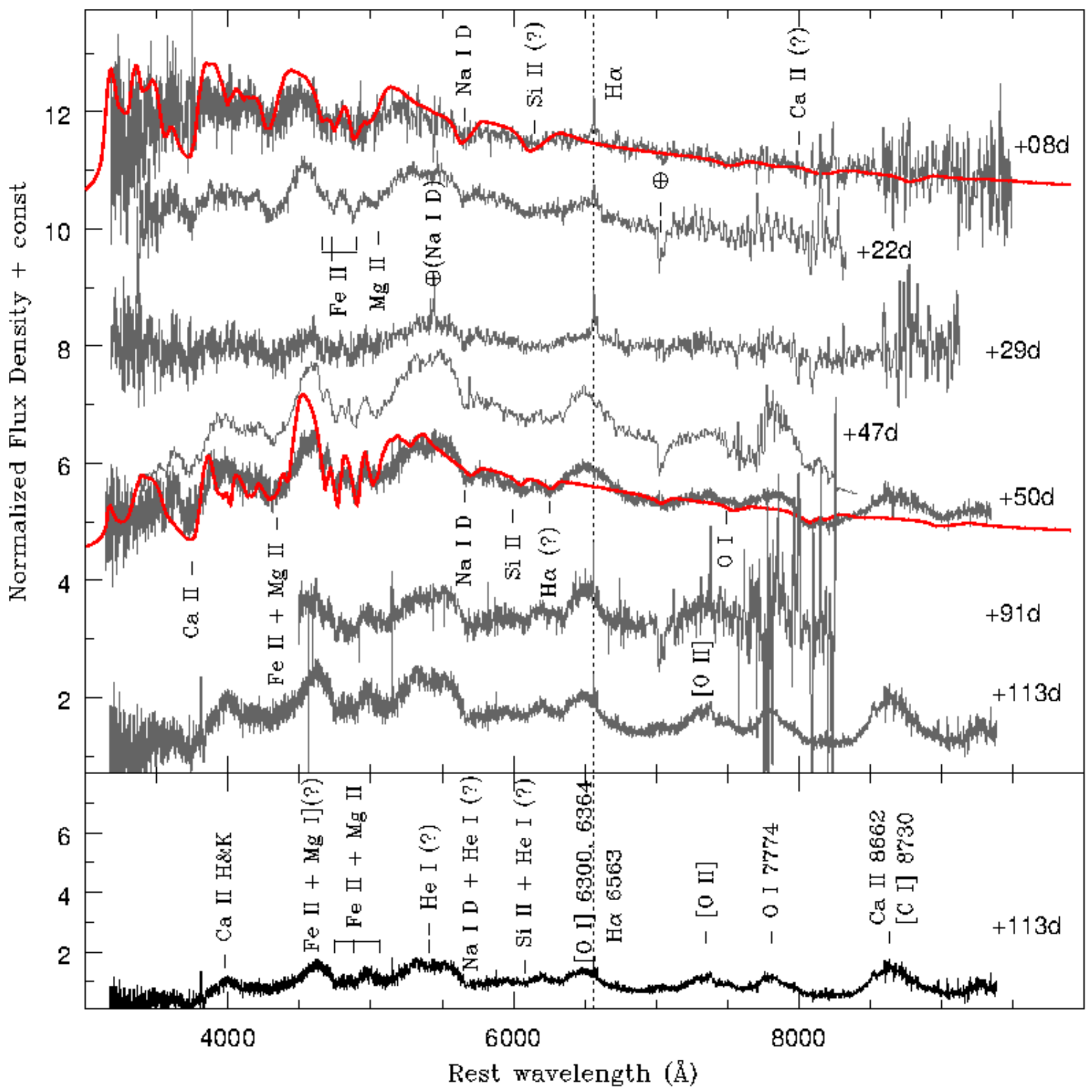}%
\caption{Spectral evolution of SN 2012aa. All spectra were
 normalized to the average continuum flux density measured in the 
 line-free region 6700--7000\,\AA\ and additive offsets were
 applied for clarity.
 The dotted vertical line shows the position of \ha. Phases are with
 respect to the date of maximum $V$-band light. The spectra obtained on 
 +8\,d and +51\,d were modeled using SYNOW (red curves) to identify some
 important line features.}
\label{fig:spectra}
\end{figure}
\section{Spectroscopic Evolution of \sn} \label{spec} 
 The upper panel of Fig. \ref{fig:spectra} presents the spectral evolution
 of \sn\ at 7 phases, starting from +8\,d to +113\,d after maximum brightness. 
 The spectra have been corrected for the redshift of the host galaxy ($z=
 0.083$; see Sect. \ref{DistExt}). In the lower panel, we show the +113\,d
 spectrum once again with some important lines marked. The line identification
 is done mainly by comparing the spectra with previously published SN~Ic and
 SLSN spectra having line identifications  \citep{2008MNRAS.383.1485V,
 2010A&A...512A..70Y,2013Natur.502..346N}. They are also based on models built
 using the fast and highly parameterized SN spectrum-synthesis code, SYNOW
 \citep{1999ApJ...527..746M,2002ApJ...566.1005B}.

 The position of narrow \ha\ is marked by a dotted vertical line. The narrow
 \ha\ in the +8, +22, +29, +91, and +113\,d spectra is from the host galaxy.
 The separation between the SN and the galaxy center is $\sim1.96$\arcsec\
 (Sect. \ref{Host}), comparable with the slit width. Thus, the detection of
 \ha\ is highly dependent on the slit orientation and seeing conditions on the
 night of observations. For example, in the observations on +47\,d and +50\,d,
 no host-galaxy line was found in the two-dimensional spectra of the SN.
 Similarly, the sharp spikes at $\sim5441$\,\AA\ in the +22\,d and
 +29\,d rest-wavelength spectra are actually the footprints of telluric emission
 from \Nai\,D vapor ($\lambda\lambda$5890, 5896) that was saturated during the
 observations. So, although the initial ($\textless$ +50\,d after maximum)
 spectra exhibit some relatively strong lines with weak broad components, 
 which may resemble the electron-scattering effect during shock interaction,
 these are most likely emission lines from the host galaxy and/or unresolved
 telluric emissions in the low-resolution spectra. 
 
\subsection{Key features}\label{spkey1}
\begin{figure*}
\centering
\includegraphics[width=8.5cm]{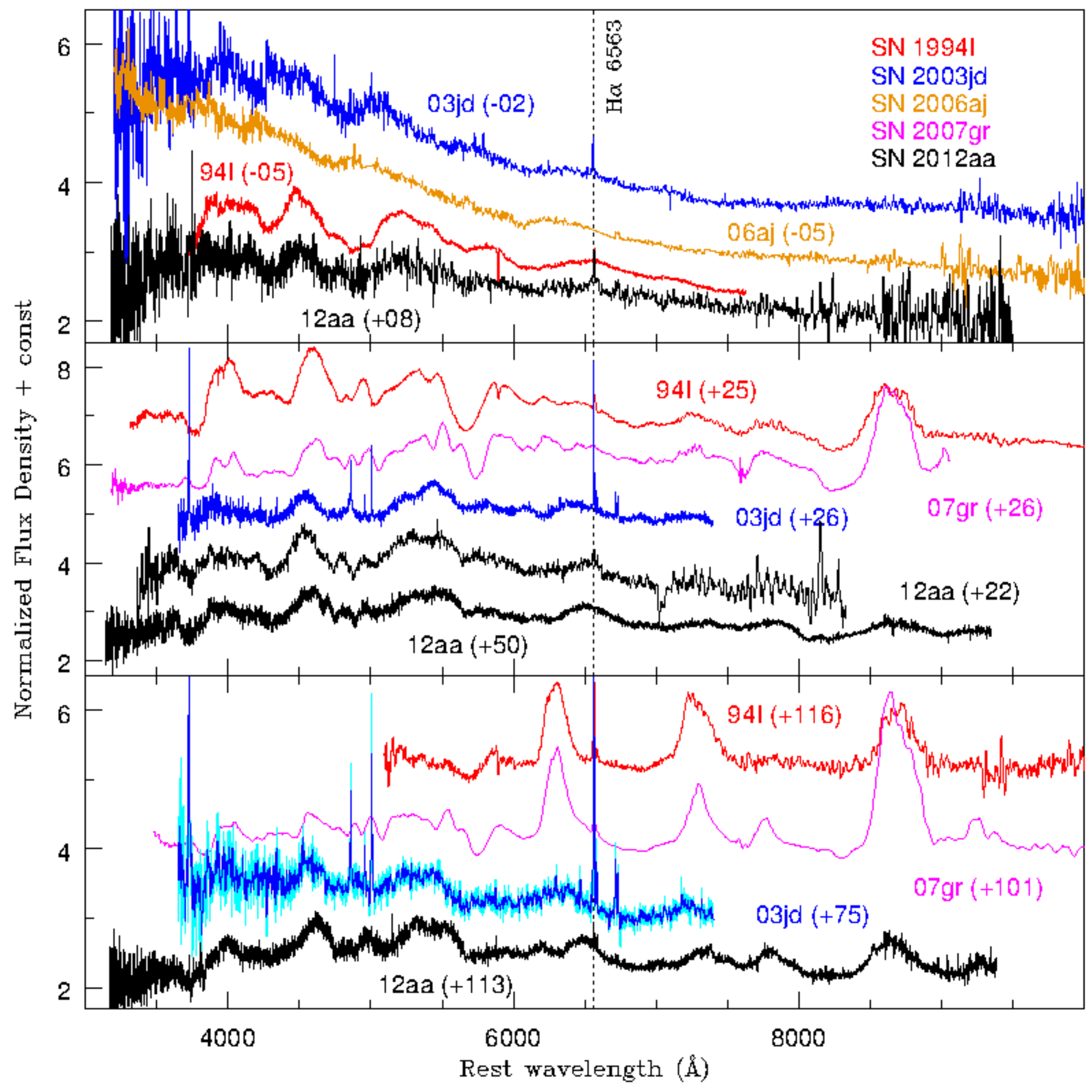}
\hskip 0.5cm
\includegraphics[width=8.5cm]{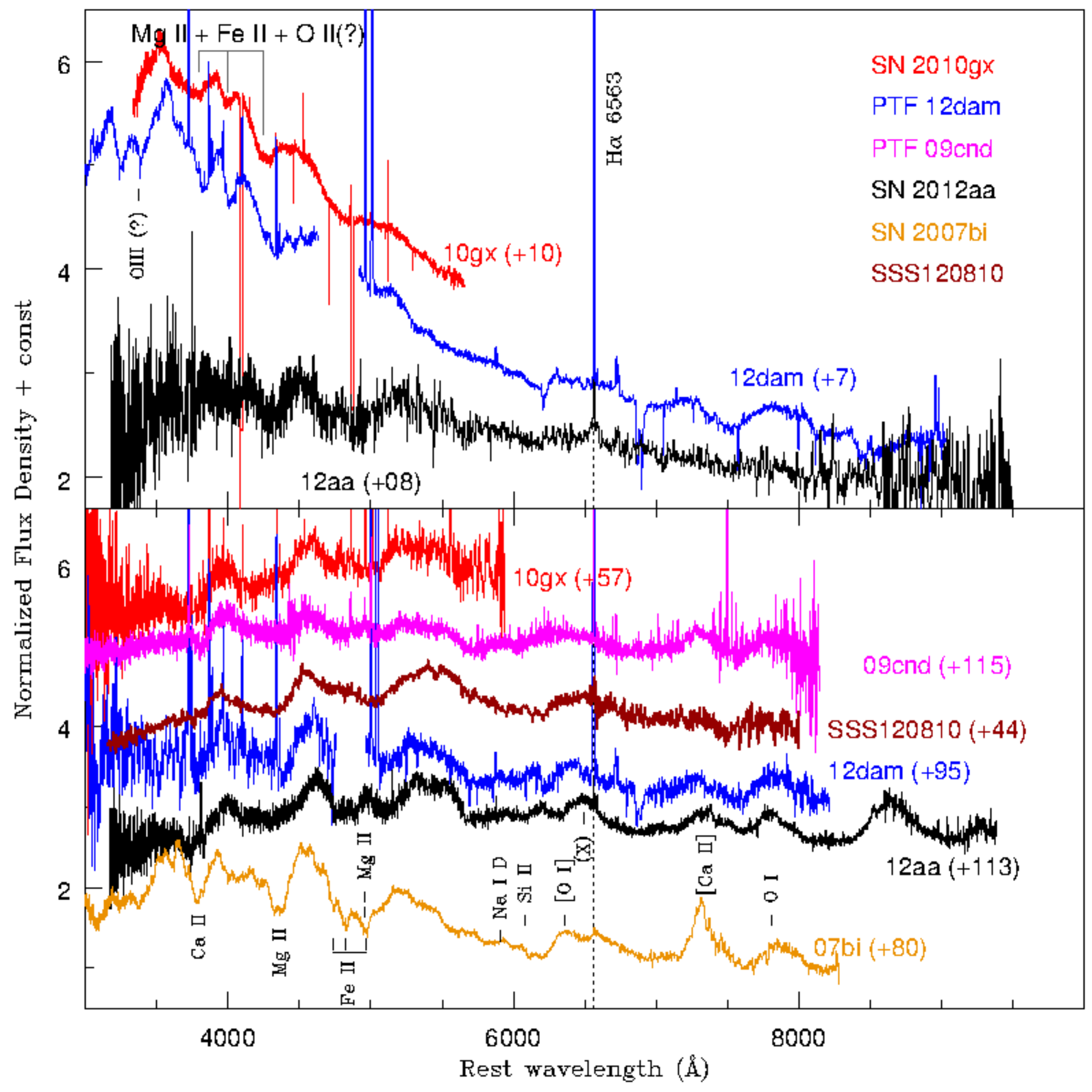}
\caption{Spectral comparison between \sn\ and other H-poor events.
 Phases are with respect to the date of maximum light. {\bf Left panel:}
 similarities between \sn\ and the broad-lined Type Ic SNe 2003jd
 \citep{2008MNRAS.383.1485V}, 2006aj \citep{2014AJ....147...99M}, and 
 2007gr \citep{2008ApJ...673L.155V}, as well as the normal Type Ic SN 1994I
 \citep{2014AJ....147...99M}, are shown. Spectra of \sn\ are similar,
 but redshifted relative, to those of SN 2003jd. However, unlike the
case for SN 2003jd, we can see \Feii\ lines in \sn\ at early stages, quite
 similar to what was observed in SNe 1994I and 2007gr. {\bf Right panel:}
 comparisons between \sn\ and Type Ic SLSNe (e.g., SN 2010gx,
 \citealt{2010ApJ...724L..16P}; SN 2007bi, \citealt{2010A&A...512A..70Y}; 
 PTF12dam, SSS120810, \citealt{2013Natur.502..346N, 2014MNRAS.444.2096N}; 
 and PTF09cnd, \citealt{2011Natur.474..487Q}) are shown. The absence 
 (or weakness) of \Oii\ and \Oiii\ lines  
 in the early-time spectra of \sn\ distinguishes it from
 canonical Type Ic SLSNe. However, near +100\,d, the spectra (except
 those of SN 2007bi) are
 similar to each other, as shown in the lower half of this panel.}
\label{fig:spcmp}
\end{figure*}

 The blackbody fit as well as SYNOW modeling of the spectra show that the
 blackbody temperature of the photosphere between +8\,d and +113\,d is between 
 roughly within $\sim8000$\,K and 7000\,K.
 The initial spectrum (+8\,d) is dominated by a continuum with some broad lines.
 With time, the SN spectra become redder and the nebular lines start to
 appear. In Fig. \ref{fig:spcmp}, a comparison of the spectral evolution
 of \sn\ and that of SNe~Ic and H-poor SLSNe is presented.

 The initial spectrum (+8\,d) of \sn\ is similar to (although redder than) 
 that of the broad-lined Type Ic SNe 2003jd and 2006aj at $-2$\,d and $-5$\,d
 (respectively) relative to maximum brightness (see the top-left panel in Fig.
 \ref{fig:spcmp}). We also compared the spectra of \sn\ taken at +8\,d and
 +50\,d after maximum with synthetic spectra generated with SYNOW. The spectra
 at late epochs are more emission-line dominated (because the ejecta became more
 optically thin), and SYNOW modeling becomes less useful. As shown in Fig.
 \ref{fig:spectra}, most of the weak, broad lines in the bluer part of the +8\,d
 spectrum are reproduced with broad \Feii\ (Gaussian FWHM $\sim 14,000$ \kms\,);
 also, the \Caii\,H\&K, \Nai\,D, and \Caii\,$\lambda$8662 absorption dips are
 present. However, unlike broad-lined SNe Ic and more like normal
 SNe~Ic (e.g., SNe 1994I and 2007gr), \sn\ starts to show narrower
 iron features relatively quickly after the epoch of maximum (
 Gaussian FWHM of \Feii\,$\lambda\lambda$4924, 5018, 5169 lines became $\sim
 5,000$ \kms\ by +22\,d).
 
 The noisy absorption feature at 6150\,\AA\ in the first spectrum can be
 reproduced either by \SiII\ at 12,500\,\kms\ or by \ha\ at 20,000\,\kms, while
 the photospheric velocity is $\sim12,000$\,\kms\ (see Sect. \ref{vphot}). This
 feature can be blended with \Feii\ lines as well.

 The +22\,d and +50\,d spectra of \sn\ are similar to the +26\,d spectrum of SN
 2003jd, with prominent features of \Caii\,H\&K, \Feii\ blended with \Mgii\
 (and maybe with \Mgi] $\lambda$4571 beyond +50\,d), as well as \Caii,
 \Oi, and \Oii\ lines. By +47\,d, \SiII\ absorption becomes more prominent.
 Finally, as shown in the bottom-left panel of Fig. \ref{fig:spcmp},
 unlike SNe 2012aa and 2003jd, beyond +100\,d the normal SNe~Ic (e.g.,
 SNe 1994I and 2007gr) become optically thin, marked by the presence of 
 strong nebular emission features of \Oia\ $\lambda\lambda$6300, 6364 
 and \Caiia.

 In contrast to \sn, the spectra of Type I SLSNe near (or before) maximum
 exhibit an additional ``W''-like absorption feature between 3500 and 
 4500\,\AA, which is the imprint of \Oii\ \citep{2010ApJ...724L..16P,
 2011Natur.474..487Q, 2013ApJ...770..128I}. As shown in the right panel of Fig.
 \ref{fig:spcmp}, the early (+8\,d) spectrum of \sn\ is redder in comparison to
 the +7\,d spectrum of PTF12dam and the +10\,d spectrum of SN 2010gx. The
 features in the blue part of the first spectrum of \sn\ are also not similar to
 those found in SN 2010gx and PTF12dam at comparable epochs. However, the 
 late-epoch ($\sim+113$\,d) spectrum of \sn\ is similar to that of SNe 2007bi,
 2010gx, PTF09cnd, and PTF12dam.

 Comparison of the spectral evolution of \sn\ with that of
 both SNe~Ic and SLSNe suggests that the broad emission feature (Gaussian
 FWHM $\approx14,000$\,\kms) near \ha\ (marked as ``X'' in the right panel of
 Fig. \ref{fig:spcmp}) could be the emergence of \Oia\ $\lambda\lambda$6300,
 6364. It is prominent in the +47\,d spectrum of \sn\ and marginally present
 already at +29\,d. Although in comparison to other known events this feature is
 highly redshifted ($\sim9500$\,\kms\ on +47\,d) in \sn, a similar
 time-dependent redshift is also noticeable for other lines in \sn\ (see Sect.
 \ref{vline}).

 Alternatively, this feature could be blueshifted ($\sim3000$\,\kms) \ha\
 emission that remains constant throughout its evolution. It becomes prominent
 by +47\,d, contemporary with the peak of the secondary light-curve bump and
 consistent with the CSM-interaction scenario. Between +47\,d and +113\,d, the
 width (FWHM) of this feature remains at around 14,000\,\kms. The values of both
 the blueshift and FWHM would be higher than the corresponding values calculated
 for CSM-interaction powered SNe (see, e.g., \citealt{2014ApJ...797..118F}).
 Moreover, we do not detect any other Balmer lines in the spectra --- though it
 is also probable that they would be highly contaminated by \Feii\ and other
 metal lines, making them difficult to detect in low-resolution spectra.

 It is noteworthy that a broad feature at similar rest wavelength was
 observed in a few other SLSNe (such as LSQ12dlf and SSS120810) during their 
 early phases and identified as the \SiII\ $\lambda6350$ doublet
 (\citealt{2014MNRAS.444.2096N}; see also Fig. \ref{fig:spcmp}). However, 
 unlike feature ``X'' (which was strong until +113\,d), the feature in LSQ12dlf
 and SSS120810 was almost washed-out by +60\,d. 

\begin{figure}
\centering
\includegraphics[width=8.5cm]{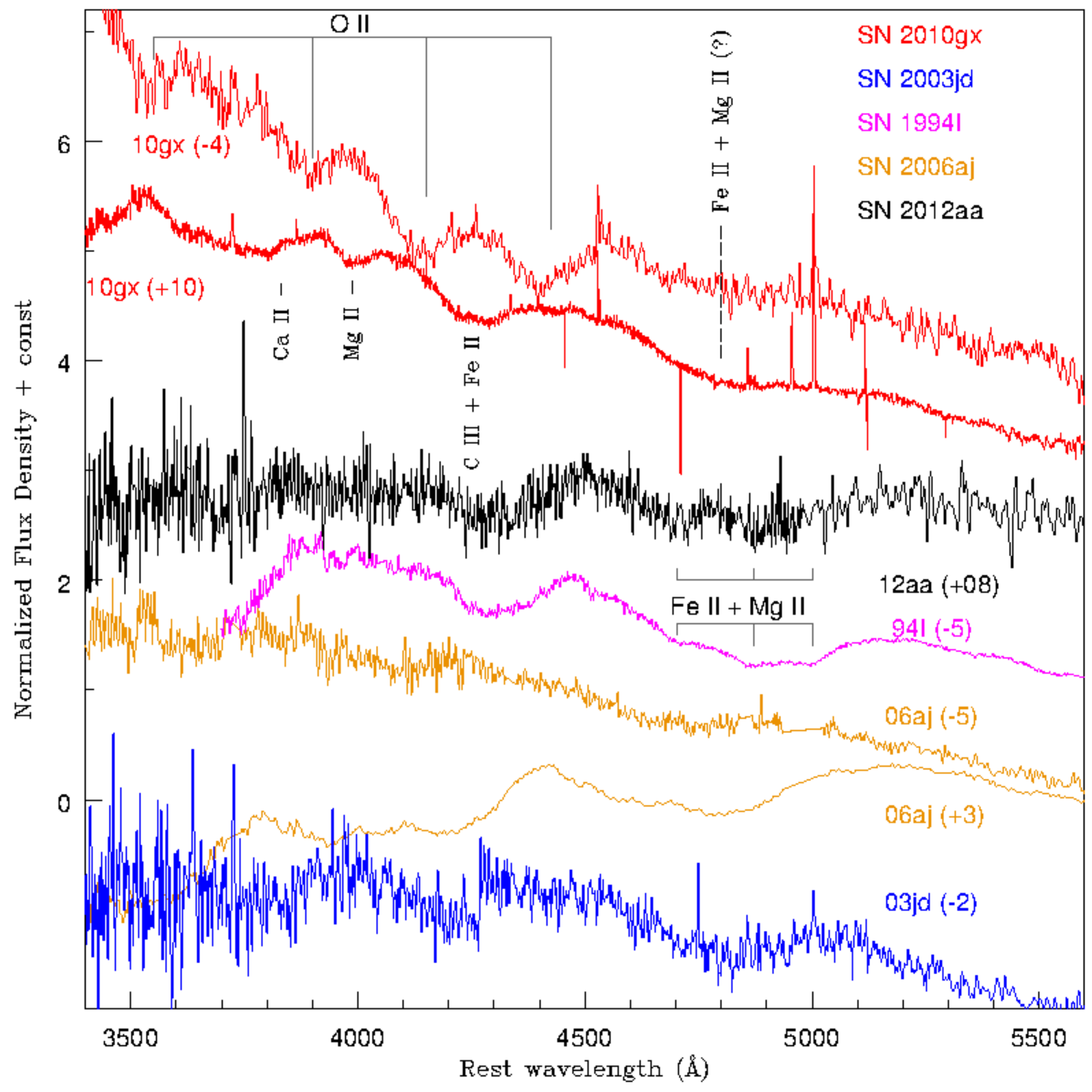}
\caption{The blue part (3400--5600\,\AA) of the early-time spectra of \sn, 
 and comparison with other SNe~Ic and H-poor SLSNe.
It is difficult to determine whether \sn\ belongs to the SLSN or CCSN 
categories.}
\label{fig:spcmp_early}
\end{figure}

 We next compare the blue part (3400--5600\,\AA) of the first spectrum of
 \sn\ with the early spectral evolution of SLSNe (e.g., SN 2010gx) and CCSNe 
 (e.g., SNe 1994I, 2003jd, 2006aj; see Fig. \ref{fig:spcmp_early}). Although the
 broad  features of \sn\ during this phase (such as \Caii, \Mgii, and \Feii\
 with Gaussian FWHM $\approx12,000$\,\kms) are similar to those in the +10\,d 
 spectrum of SN 2010gx, the early key features of SLSNe (like \Oii\ lines) are
 not detected in the spectra of \sn. The presence of the \Feii\ 
 and \Mgii\ blend (Gaussian FWHM $\approx150$\,\AA\,) around 4800\,\AA\ in
 this early-time spectrum distinguishes it from canonical
 SLSNe and broad-lined CCSNe (like SNe 2003jd and 2006aj). At comparable epochs,
 both SLSNe and broad-lined SN~Ic spectra show a very
 broad absorption dip (FWHM $\approx350$\,\AA\ in SN 2010gx), which is probably
 a blend of \Feii\ multiplets. On the other hand, the Type Ic SNe 1994I and 
 2007gr exhibit \Feii\ and \Mgii\ blends that are quite similar to those of \sn.

 Furthermore, in terms of the timescales, the features 
 that distinguish \sn\ from SLSNe are visible in the red part
 ($\textgreater$ 7000\,\AA) of the spectra. The lines
 \Caiia\ $\lambda\lambda$7291, 7324, \Oiia\ $\lambda\lambda$7319, 7330,
 \Oi\ $\lambda7774$, \Caii\ $\lambda\lambda$8498, 8542, 8662, and \Cia\
 $\lambda8730$, which are strong in \sn\ spectra starting from +50\,d, are
 barely detected before +80\,d in the SLSN samples presented by
 \citet{2013ApJ...770..128I}. \citet{2015MNRAS.452.3869N} also noticed that,
 although the \Oi\ and \Caii\ near-infrared lines start appearing beyond
 +50\,d in SLSNe, these features (along with other forbidden lines like \Caiia)
 become strong only at $\gtrsim+100$\,d after maximum. These lines are instead
 common in SNe~Ibc, even just after maximum brightness (e.g.,
 \citealt{1999ApJ...527..746M, 2002ApJ...566.1005B, 2008MNRAS.383.1485V,
 2013MNRAS.434.2032R}), and also in SNe~II during the late photospheric phase
 \citep{1994AJ....108.2220F, 2004MNRAS.347...74P}. In this sense, \sn\ is more
 similar to SNe~Ic than to Type I SLSNe. 

 To summarize, although the overall spectral evolution of Type I SLSNe is
 similar to that of canonical SNe~Ic, the timescales for SLSNe are much longer
 than for their corresponding low-luminosity versions.
 The nature of the \sn\ light curve is more similar to that of SLSNe, while the
 timescale (and nature) of its spectral evolution more closely resembles 
 that of normal SNe~Ic.

 However, there are caveats. The Type I SLSNe like PTF11rks started to exhibit
 \Feii, \Mgii, and \SiII\ features from $\sim +3$\,d post maximum, as is 
 common in SNe~Ic. Beyond +60\,d the distinct \Feii\ multiplet features near
 5000\,\AA\ were noticed in SLSN PTF11hgi \citep{2013ApJ...770..128I,
 2015MNRAS.452.3869N}. These were not prominent in the broad-lined SNe~Ic 
 associated with engine-driven explosions such as SNe 1998bw,
 2006aj, 2003dh, 2003jd, and 2003lw at comparable epochs, although present in
 the Type Ic SNe 1994I and 2007gr, and also in stripped-envelope SNe~Ib like
 SN 2009jf \citep{2011MNRAS.416.3138V}. It is also worth noting that the
 emission lines in \sn\ start appearing at a relatively early stage in
 comparison to what is seen for SLSNe, and there is not a considerable change in
 the features beyond +50\,d. Almost all features persist up to the epoch of the
 last (+113\,d) spectrum with comparable line ratios, which is unusual in normal
 SNe~Ic. In that sense, the timescale of \sn\ is longer than that of most normal
 SNe~Ic.

\subsection{Photospheric velocity}\label{vphot}
\begin{figure}
\centering
\includegraphics[width=8.5cm]{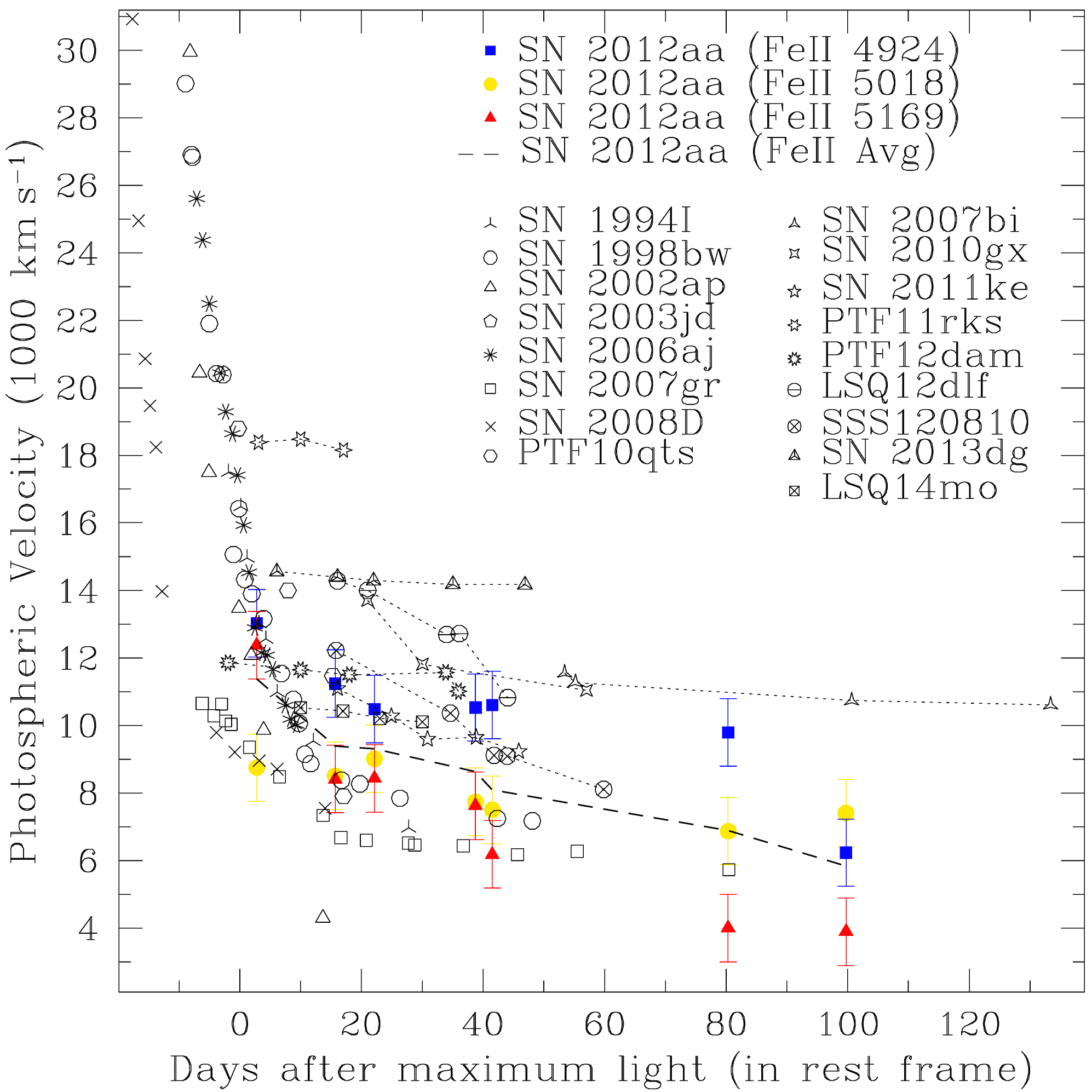}%
\caption{Comparison of the photospheric velocity of SN 2012aa with that of
 SNe~Ibc and SLSNe. Velocities of \Feii\ $\lambda\lambda$4924, 5018, 5169 
 absorption in \sn\ are indicated with filled squares, circles, and triangles,
 respectively. The dashed line represents the average photospheric
 velocity of \sn\ calculated from the \Feii\ lines. All of the 
 transients labeled
 in the left column are SNe~Ibc, while the SLSNe~Ic are labeled in the
 right column. For all SLSNe, the velocities at individual epochs have been
 connected with dotted lines. The photospheric velocities (measured from
 \Feii\ and \SiII\ lines) of other transients 
 in the plot have been adopted from the
 literature: SN 2007bi from \citet{2010A&A...512A..70Y}; all other
 SLSNe are from \citet{2015MNRAS.452.3869N}; SNe 2002ap, 2006aj, and 2008D are
 from \citet{2008Sci...321.1185M}; SNe 1994I and 1998bw are from
 \citet{2001ApJ...555..900P}; SNe 2003jd, 2007gr, and PTF10qts are
 from \citet{2008MNRAS.383.1485V}, \citet{2009A&A...508..371H}, and
 \citet{2014MNRAS.442.2768W}, respectively.}
\label{fig:photvel}
\end{figure}

 The \Feii\ $\lambda\lambda$4924, 5018, 5169
 absorption lines throughout the spectral evolution helped us determine the
 photospheric velocity ($v_{\rm phot}$) of \sn.
 Figure \ref{fig:photvel} illustrates a comparison of the velocities of these
 metal lines in \sn\ with the photospheric velocities of other SNe~Ic and
 SLSNe. Uncertainties are calculated by estimating the scatter in repeated
 measurement of the absorption dips; the error bars represent the 2$\sigma$
 variation. Although near peak brightness the velocities of
 individual lines are comparable with each other, at late stages a deviation in
 their velocites is noticed. The dashed line represents the average
 velocities of these three lines and can be considered as representative of the
 photospheric velocity of \sn.

 The average evolution of these Fe lines sets $v_{\rm phot}$ to
 $\sim11,400$\,\kms\ at maximum brightness. The comparison (Fig.
 \ref{fig:photvel}) shows the \sn\ velocity
 evolution is comparable to that of SNe~Ibc rather than to the shallow velocity
 profiles of SLSNe. The flattened velocity profiles of SLSNe and in contrast
 to that for CCSNe has been seen in moderate samples ($\sim15$ objects,
 \citealt{2011ApJ...743..114C, 2015MNRAS.452.3869N}; see also Fig
 \ref{fig:photvel}). Such a velocity evolution in SLSNe had been predicted as a
 consequence of the magnetar-driven explosion model; if the energy released by
 the spin-down magnetar is higher than the kinetic energy of the SN, the entire
 ejecta will be swept up in a dense shell and will propagate with a uniform
 velocity until the shell becomes optically thin to electron scattering on a
 timescale of 100\,d after explosion \citep{2010ApJ...717..245K}. On the other
 hand, a power-law decline of the photospheric velocities in SNe~Ibc is a
 consequence of homologous expansion. Since the average photospheric velocity
 profile of \sn\ is similar to that of CCSNe, a homologous expansion is 
 more likely in \sn\ than a magnetar-driven scenario.

\subsection{Line velocities}\label{vline}
\begin{figure}
\centering
\includegraphics[width=8.5cm]{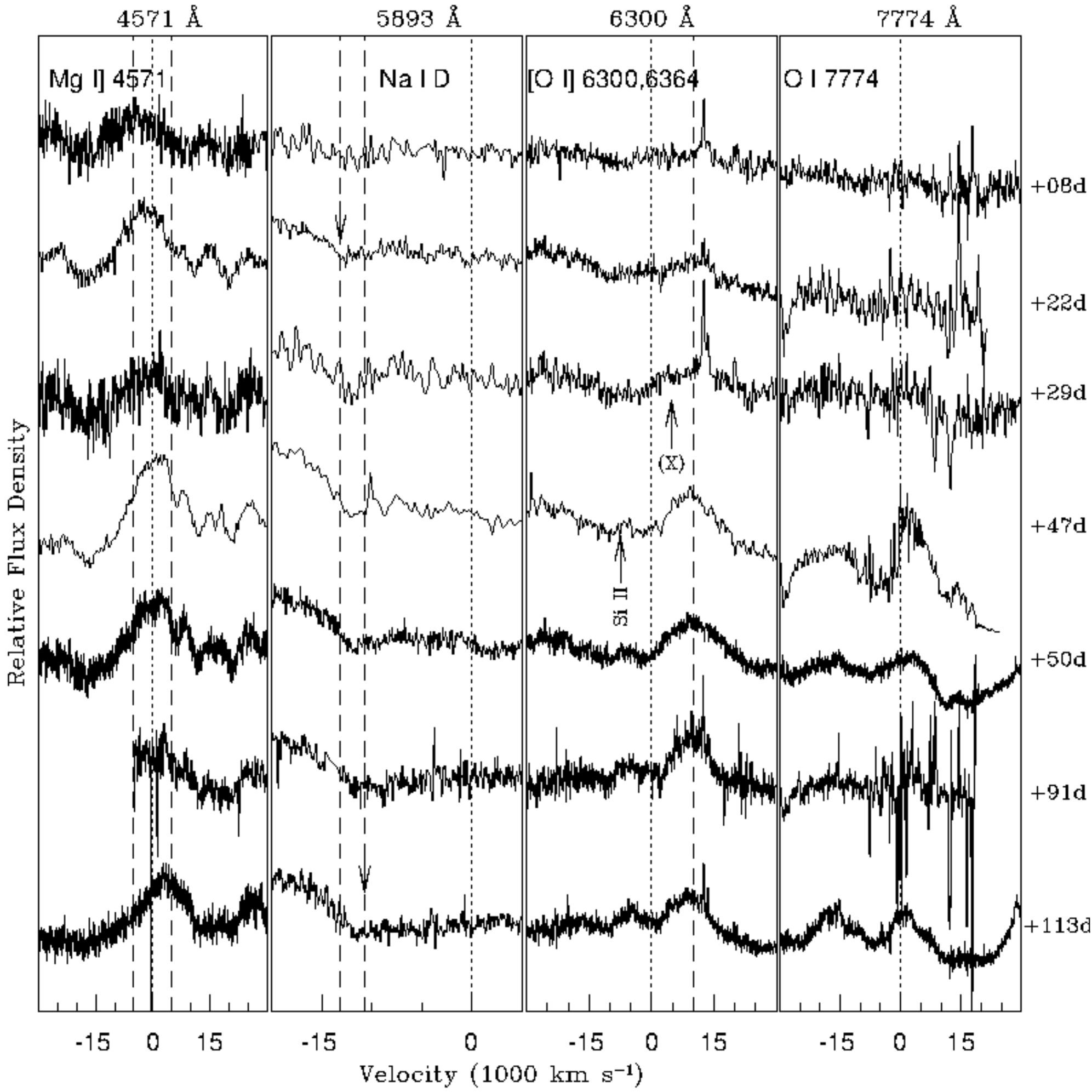}%
\caption{Temporal evolution of some spectral windows in \sn. The zero velocity
 shown with a dotted line in each panel marks the rest wavelength of the
 corresponding element mentioned at the top of the panels. The flux density 
 scale is relative. The flux of each line has been scaled in the same
 way as for the entire spectra mentioned in Figure \ref{fig:spectra}.
 The dashed lines are the boundaries between which the emission peaks and
 absorption dips get shifted during the SN evolution. The downward arrows
 in the second panel show the blueshift of the \Nai\,D absorption.
 The arrow marked with ``X'' shows the emergence of \Oia\ by +29\,d.}
\label{fig:linevel}
\end{figure}

 Figure \ref{fig:linevel} shows the velocity evolution of different lines at
 around 4571\,\AA, 5893\,\AA, 6300\,\AA, and 7774\,\AA. The zero
 velocity of each panel corresponds to the aforementioned rest wavelengths
 (marked at the top of each panel).

 If the SN emission is not shock-dominated, the spherical homologous expansion
 of optically thick ejecta would produce a spectrum with simple P-Cygni
 line profiles that are a combination of emission and absorption features;
 the emission peak is always centered at the rest wavelength of the line
 (and hence at zero velocity in the rest frame), while the absorption dip 
 gets blueshifted, which is a measure of
 outflow velocity. Unlike this simple P-Cygni feature, we noticed a
 velocity evolution of the emission components of several spectral lines in
 \sn. As shown in Figure \ref{fig:linevel}, 
 near maximum light the emission feature (tentatively marked as a blend of
 \Feii\ and \Mgii) near 4571\,\AA\ was blueshifted from the rest position by
 $-5000$\,\kms. With time this feature became redshifted by 10,000\,\kms, and
 finally at +113\,d it was redder with a projected velocity of
 $\sim+$5000\,\kms. Similarly, after its emergence at around +29\,d, the \Oia\
 $\lambda\lambda$6300, 6364 doublet (marked by the ``X'' in the third panel)
 also exhibits a redshift of 10,000\,\kms, consistent with the redshift of the
 emission feature at 4571\,\AA. Similar velocity evolution of the emission lines
 could indicate an aspherical evolution of these line-forming regions. 
 This kind of analysis of course depends on the assumption that we can identify
 single elements that dominate the time evolution of their spectral region. The
 redshift of the
 feature near 4571\,\AA\ may be caused by the emergence of the \Mgia\ emission
 line. However, in SNe~Ibc at comparable epochs ($\gtrsim110$d post maximum) the
 strength of this semi-forbidden line is often much less than that of other
 nebular lines (e.g., \citealt{2009MNRAS.397..677T}, their figure 1). In
 \sn\ at similar epochs this feature is stronger than that observed in SNe~Ibc.
 Moreover, $\sim5,000$\,\kms\ redshift with respect to its rest position at
 this stage (+113d) of evolution makes this line quite unlikely to be \Mgia\ in
 a spherically expanding SN envelope.

 The absorption dip of the \Feii\ plus \Mgii\ blend is at $\sim-18,200$\,\kms\
 in the first spectrum, while in the last spectrum it is at $\sim-13,200$\kms.
 Similarly, the FWHM of \Oia\ at +47\,d is $\sim11,900$\,\kms, and it
 decreased to $\sim9700$\,\kms\ by +113\,d.
 Similarly the absorption dip of the \Nai\,D line is at $\sim13,000$\,\kms\
 at +22d and decreased to $\sim10,500$\,\kms\ by +113d post maximum.
 Such values are higher than the photospheric velocity (Sect. \ref{vphot}),
 indicating that these line-emitting regions could be detached from the
 photosphere. 

 The shapes of line profiles of Fe-group elements and \Oia\ are dependent on the
 distribution of $^{56}$Ni inside the ejecta and/or can be affected by
 blending (even for a symmetric distribution of $^{56}$Ni). Even with an
 extremely aspherical distribution of $^{56}$Ni inside spherical ejecta, the
 \Oia\ profile can be distorted by $\sim 3000$\,\kms\
 \citep{2008Sci...319.1220M, 2008ApJ...687L...9M}. Thus, the
 highly redshifted lines in \sn\ probably indicate an extreme asphericity in the
 ejecta themselves --- perhaps the ejection of a blob receding away at
 $\sim10,000$\,\kms.

 The \Oi\ $\lambda$7774 line, which started to emerge almost simultaneously with
 \Oia\ $\lambda\lambda$6300, 6364 (see fourth panel), evolved like a canonical
 P-Cygni feature: the emission peak remained at around zero velocity, while the
 absorption dip was initially blueshifted by $\sim-10,000$\,\kms, slowed down,
 and finally stayed at $\sim-5000$\,\kms. This suggests a spherical homologous
 expansion of this particular line-emitting region and
 contrasts with the evolution of the ``X'' profile, if we assume it to be
 \Oia\ $\lambda\lambda$6300, 6364. We speculate that this situation can
 only happen under extreme
 asymmetry, when a blob enriched with \Oia\ $\lambda\lambda$6300, 6364 is
 ejected away from the expanding SN with a velocity of $\sim10,000$\,\kms.

\begin{figure}
\centering
\includegraphics[width=8.5cm]{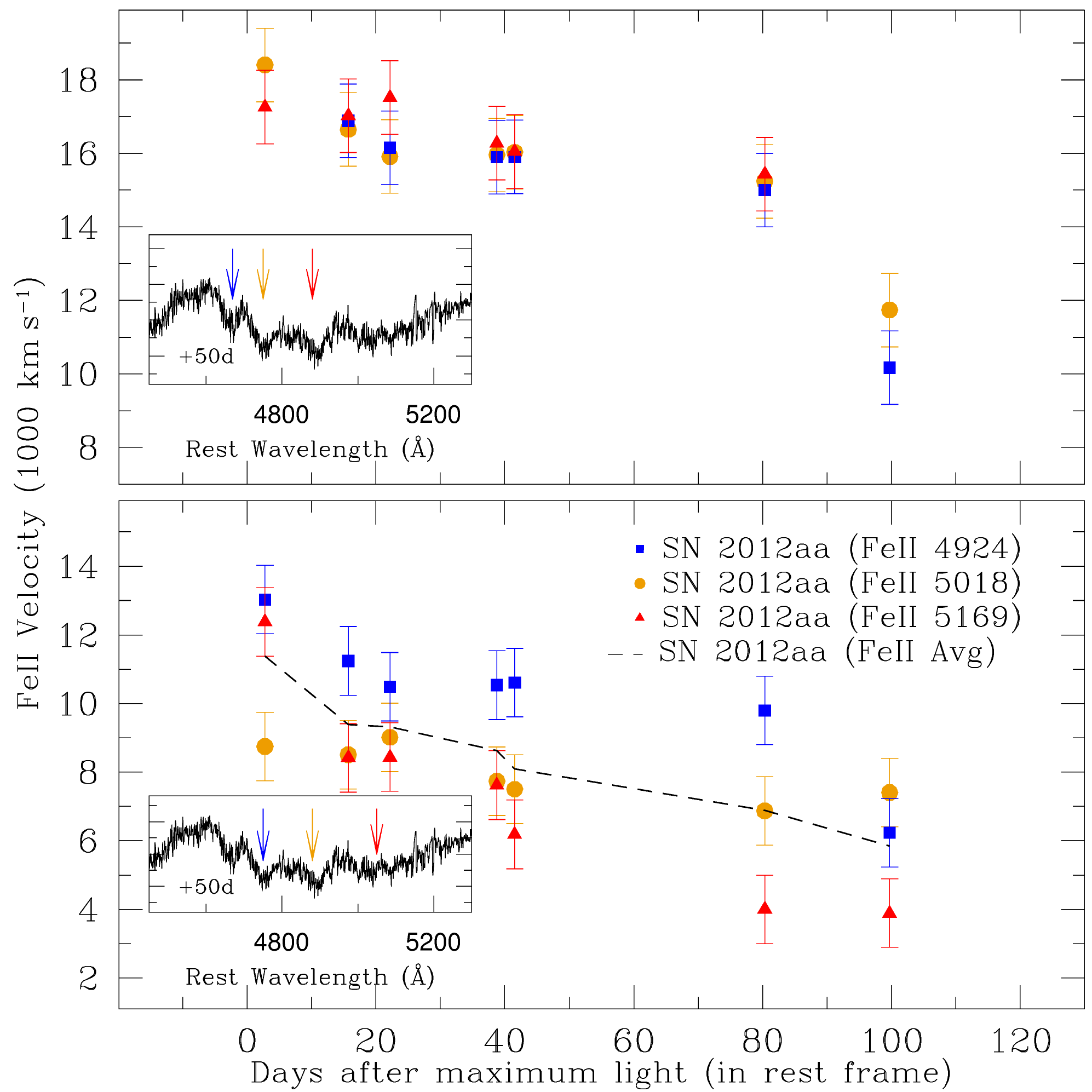}%
\caption{The velocity evolution of \Feii\ lines in SN 2012aa. The three 
 lines \Feii\ $\lambda\lambda$4924, 5018, 5169 are difficult to identify
 unambiguously. We found two different possibilities in \sn, shown here
 using the +50\,d spectrum; the lines are respectively marked with blue, 
 orange, and red arrows in the insets. {\bf Lower panel:} the velocity 
 evolution is a power law common in CCSNe. {\bf Upper panel:}
 the velocity evolution has a shallow decline common in SLSNe.}
\label{fig:FeIIvel}
\end{figure}

\subsection{\Feii\ velocities}\label{Feline}
 In Fig. \ref{fig:FeIIvel}, we show the +50\,d spectrum of \sn, marking three
 absorption lines of \Feii\ $\lambda$4924, $\lambda$5018, and $\lambda$5169
 in blue, orange, and red (respectively). The lower panel illustrates 
 that in one set of combinations the \Feii\ velocity profile will be
 close to a  power law, representing homologous expansion and comparable to the
 photospheric velocity profiles of CCSNe. This velocity is consistent with the
 profile of other elements, and we therefore consider it as the representation
 of the photospheric velocity of \sn\ (used in Sect. \ref{vphot} and Sect.
 \ref{vline}). On the other hand, in a different set of absorption-line
 combinations (shown in the upper panel of the figure), we would get a shallow
 velocity profile, more consistent with that of SLSNe (e.g., see Fig.
 \ref{fig:photvel}). 

 Although the power-law profile is more feasible in \sn, the shallow decline
 cannot be completely ruled out. In this case the velocity profiles of
 individual \Feii\
 lines are actually more mutually consistent, and in all SYNOW
 models we get a better fit for \Feii\ lines by keeping their velocities
 higher than the photospheric velocity, and evolving them as a detached shell.
 From the spectrum (upper panel, Fig. \ref{fig:FeIIvel}) we can also see 
 that Gaussian FWHM values ($\sim3000$\,\kms) of the \Feii\ lines in this
 combination are
 more consistent with each other than the alternative combination (lower panel).
 If the shallow profile (upper panel) would be maintained by the \Feii\ in
 \sn, along with a lower photospheric velocity profile (like in the lower 
 panel), we may draw the following conclusions.

 The velocity of \Feii\ lines at maximum brightness is $\sim18,000$\,\kms,
 declines to $\sim16,000$\,\kms\ within 20 days, and remains constant at 
 that value for the next 60 days, finally dropping to 12,000\,\kms\ at 
 100\,d after the peak. In this scenario, the velocity profile of 
 \Feii\ lines in \sn\ is much higher than the typical velocities of SLSNe
 (e.g., $10,500 \pm 3100$\,\kms; \citealt{2015MNRAS.452.3869N}) measured at a
 comparable phase. Higher bulk velocity
 along with a narrow velocity spread (FWHM) of \Feii\ absorption lines is
 not common among SNe~Ic; high velocities of the lines generally make the
 corresponding profiles broader \citep{2015arXiv150907124M}. 

\section{Properties of the host galaxy} \label{Host}
 The spectral and photometric differences of \sn\ compared with the 
 well-known canonical CCSNe and Type I SLSNe motivate us to 
 investigate the nature of its host.

\begin{figure}
\centering
\includegraphics[width=8.5cm]{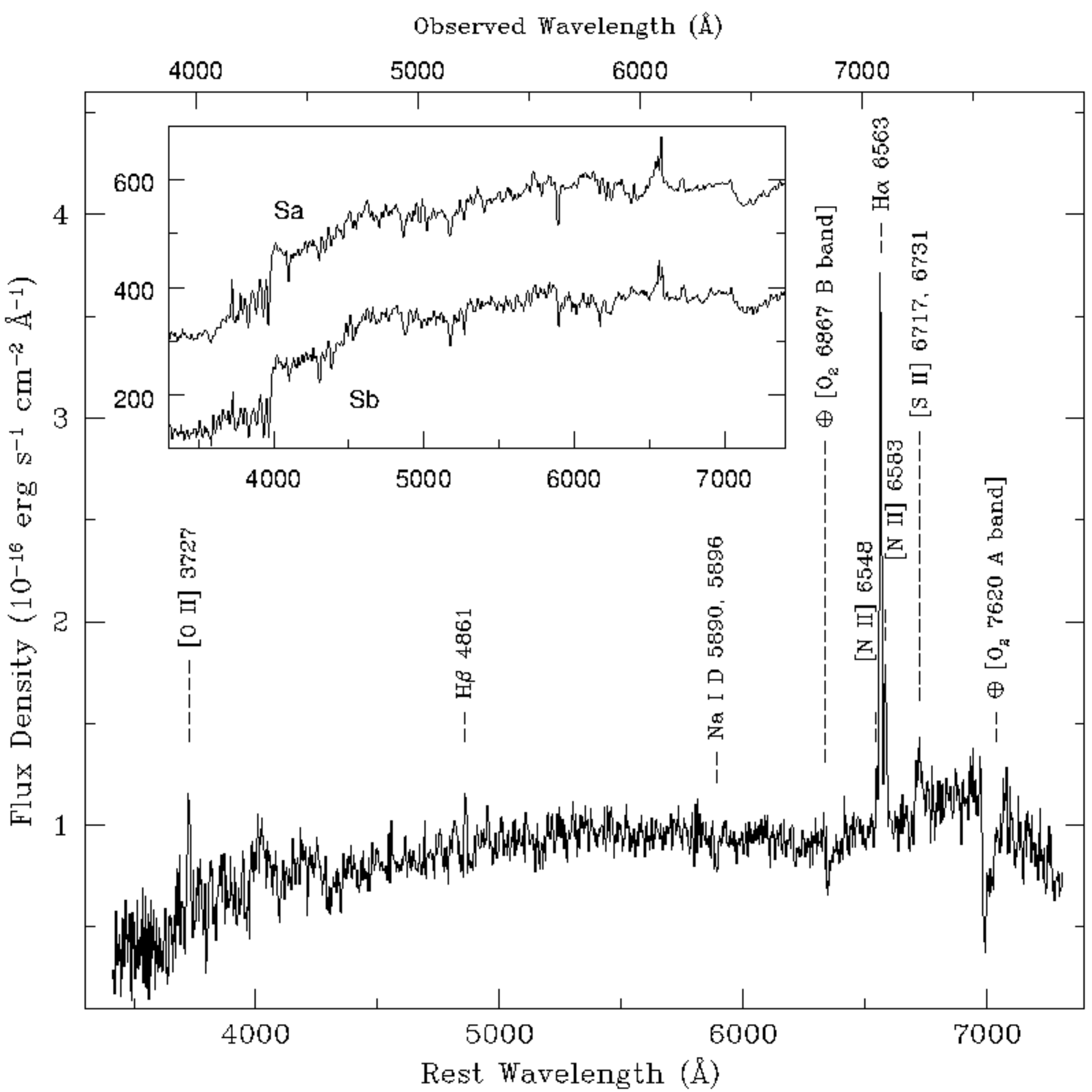}%
\caption{Optical spectrum of the host-galaxy 
 center taken with NOT +905\,d after discovery
 of SN 2012aa. Prominent lines are marked. The spectrum suggests that the
 host is a typical star-forming galaxy of Type Sa/Sb/Sbc. From the
 observed wavelength of \ha, we measured the host-galaxy redshift 
 to be $z = 0.083$.}
\label{fig:Hostspec}
\end{figure}

 Figure \ref{fig:Hostspec} presents the spectrum of the host taken $\sim 905$\,d
 after the SN discovery, when the SN had disappeared. It exhibits 
 the \ha\ line along with \Niia\ $\lambda\lambda$6548, 6584,
 \Siia\ $\lambda\lambda$6716, 6731, \hb, and \Oiia\ $\lambda$3727,
 characteristic of a spiral galaxy with Sa/Sb/Sbc subclasses
 \citep{1992ApJS...79..255K}. Fixing the morphological index (T-type)
 parameter for this galaxy as 3.3 from the ``Third Reference Catalog (RC3) of
 Bright Galaxies'' \citep{1991rc3..book.....D} and assuming (from optical
 images) a position angle of $\sim 0$\degree, we calculate a typical deprojected
 dimension of the host as $\sim6$\arcsec, corresponding to a physical size of
 $\sim8$\,kpc. Thus, the host of \sn\ is much smaller than our own Milky Way
 Galaxy and other spiral hosts of nearby SNe~Ibc.

 Measurements of the overall metallicity of the host and that at the SN
 location are important for understanding the properties of the birth places of
 SN progenitors. The first entity can be measured either in an indirect way
 from the $B$-band luminosity of the host \citep{2004ApJ...613..898T} or
 directly after calculating the \Niia\ $\lambda6584$/\ha\ ratio in the host
 nucleus \citep{2004MNRAS.348L..59P}. These diagnostics have been used by
 several similar studies (e.g., \citealt{2010MNRAS.407.2660A,
 2013A&A...558A.143T}, and references therein).

 The apparent $B$ magnitude of the host is
 $20.17\pm0.06$, which corresponds to an absolute magnitude of $\sim-18.13$.
 Using the empirical relationship between $B$-band luminosity and metallicity
 found for SDSS galaxies \citep{2004ApJ...613..898T}, we find the overall
 metallicity of the host to be [12 + log(O/H)] $={8.58}^{+0.05}_{-0.02}$\,dex.
 This implies the overall metallicity of the host, $Z\approx0.78$\,\zsun, where
 [12 + log(O/H)]$_\odot\approx8.69$\,dex is the
 solar metallicity in terms of oxygen abundance \citep{2009ARA&A..47..481A}.

 The second method is to calculate the metallicity from the \Niia\
 $\lambda6584$/\ha\ ratio,
 the so-called N2 diagnostic \citep{2004MNRAS.348L..59P}. Owing to the similar
 wavelengths of these lines, this method is neither dependent on the
 correct measurement of extinction nor affected by differential slit losses.
 From our +905\,d spectrum, we measure this flux ratio
 to be 0.391, which gives the metallicity of the host to be [12 + log(O/H)]
 $\approx8.72$\,dex, which corresponds to $Z\approx1.07$\,\zsun.

 Hence, the overall metallicity of the host is $Z_{\rm host} =
 0.92\pm0.34$\,\zsun, which is consistent both with the solar metallicity and
 with the average metallicity of the star-forming galaxies hosting SNe~IIP 
 and normal SNe~Ibc \citep{2010MNRAS.407.2660A}.

 The diagnostic tool known as a BPT diagram \citep{1981PASP...93....5B}
 is an important means for determining the nature of the host. It
 shows the distribution of galaxies in the log(\Niia/\ha) vs.
 log(\Oiiia/\hb) plane. Different schemes have been proposed to classify the
 galaxies using SDSS data \citep[e.g.,][]{2006MNRAS.372..961K}. Using
 this tool, it has been found that SLSNe generally reside in galaxies that
 are considerably different from star-forming galaxies hosting CCSNe (see Sect.
 \ref{res:comp}). We use the BPT diagram to diagnose the host 
 of \sn. We measure log(\Niia/\ha) $= -0.407$, but \Oiiia\ emission is
 absent in the host spectrum. Considering
 the continuum flux level as the minimum possible flux for of \Oiiia\, we
 find log(\Oiiia/\hb) $\lesssim-0.796$. These two quantities satisfy 
 the inequality (see \citealt{2006MNRAS.372..961K})
 log(\Oiiia/\hb) $<$ 0.61/[log(\Niia/\ha) $- 0.05] + 1.3$,  
 which defines the region of star-forming galaxies in the BPT diagram. 
 Thus, we conclude
 that the host of \sn\ is a normal star-forming galaxy. This
 is consistent with the conclusion drawn from the 
 metallicity measurement. 
 
\section{Physical parameters} \label{Progenitor}
 The SN can be powered either entirely by radioactive $^{56}$Ni $\rightarrow$
 $^{56}$Co decay or by CSM interaction (or both). Here we discuss the above two
 scenarios for \sn.

\subsection{Radioactive powering}\label{radioctivity}
 We adopt the analytical relations of \citet{1982ApJ...253..785A} to estimate
 explosion parameters from the observables. Assuming the power source to
 be concentrated in the inner parts of the ejecta, the total ejected mass
 ($M_{\rm ej}$) and the amount of $^{56}$Ni produced during explosive
 nucleosynthesis can be derived from both the peak and tail luminosities of
 \sn. It can also be measured by fitting the light curve of \sn\ with the
 stretched and scaled light curve of SN 1998bw, for which the Ni mass
 and $M_{\rm ej}$ are known \citep{2013MNRAS.434.1098C}. 

 The \bvri\ quasi-bolometric light curve of \sn\ shows a declining tail with
 a rate of 0.02\,mag\,d$^{-1}$ from +60\,d to +100\,d. This is comparable to 
 the decline rates of SNe~Ibc (see Sect. \ref{phot_late}). Under the assumption
 that for \sn\ this phase is mainly
 powered by radioactivity, the scaling relation to SN 1987A can be applied to
 estimate the amount of $^{56}{\rm Ni}$. Assuming the rise time
 of \sn\ is  $\gtrsim$30\,d, we can say that roughly 130 d after explosion, the
 luminosity of \sn\ is $(L_{\rm 12aa})_{130} \approx 2.4\times10^{42}$ \ergs.
 At a comparable phase, the \bvri\ luminosity of the well-studied Type II SN
 1987A was $(L_{\rm 87A})_{130} \approx 1.2\times10^{41}$ \ergs. Assuming that
 in both events the $\gamma$-rays from radioactive yields were fully trapped
 inside the ejecta at 130\,d after the explosion, we can write
 $(L_{\rm 12aa})_{130}/(L_{\rm 87A})_{130} \approx
 (M_{\rm Ni})_{\rm 12aa}/(M_{\rm Ni})_{\rm 87A}$, and under above mentioed
 assumption we obtain a lower-limit of the amount of radioactive Ni for \sn,
 $(M_{\rm Ni})_{\rm 12aa}\approx1.4$\,\msun, where
 $(M_{\rm Ni})_{\rm 87A} = 0.075$\,\msun\ \citep{2003A&A...404.1077E}.
 
 Following \citet{1982ApJ...253..785A}, the amount of $^{56}{\rm Ni}$ can also
 be calculated from the peak luminosity after accounting 
 for the peak width of the
 light curve, which is an estimate of the diffusion timescale. Using the same
 analytical approach, and taking only the rise time into account,
 \citet{2005A&A...431..423S} found the following relation for Type I events: 
\[ M_{\rm Ni}/{{\rm M}_\odot} = (L_{\rm peak}/{\rm erg}\,{\rm s}^{-1})/\]
\[[6.45\times10^{43}\,e^{-t_r/8.8} + 1.45\times10^{43}\,e^{-t_r/111.3}],\]
\noindent
 where $t_r$ is the rise time in days, $M_{\rm Ni}$ is in solar mass units, and
 $L_{\rm peak}$ is the peak luminosity in \ergs. However, this expression is
 more
 applicable for those cases where the rise time is comparable to the decline
 time and hence the diffusion timescale can be quantified only by $t_r$. For
 different rise and decline times, $t_r$ can be replaced by the geometric mean
 ($\tau_m$) of $\tau_{\rm ris}$ and $\tau_{\rm dec}$. For the quasi-bolometric
 curves, for \sn\ we get
 $\tau_m \approx 44$\,d. Since $L_{\rm peak} \approx 1.6\times10^{43}$\,
 \ergs, this implies $M_{\rm Ni} \approx 1.6$\,\msun, consistent with the value
 found from the late-time light curve.

 Moreover, considering both diffusion timescale and velocity at the peak the
 amount of ejected mass can be derived
 \citep{1996snai.book.....A, 2015MNRAS.452.3869N}:
\[ M_{\rm ej}\approx 0.77\, {\rm M}_\odot{(k/0.1\,{\rm cm}^2\,{\rm g}^{-1})}^{-1}(v_{\rm phot}/10^9\,{\rm cm}\,{\rm s}^{-1}){(\tau_m/10\,{\rm d})}^2,\]
\noindent
 where $v_{\rm phot}$ is the peak photospheric velocity in ${\rm cm\,s}^{-1}$
 and $k$ is the opacity. In the literature different values of $k$ have been
 adopted, normally varying between 0.05 and 0.1. To maintain consistency with
 the approach of \citet{2013MNRAS.434.1098C}, we adopt
 $k=0.07\,{\rm cm^2\,g^{-1}}$. For $\tau_m \approx 44$ days and
 $v_{\rm phot}=11,400$\,kms, we then get $M_{\rm ej} \approx 24$\,\msun.

 Here we caution the reader that this is the maximum possible 
 amount of $^{56}$Ni
 produced by \sn\ if and only if the entire
 light curve is powered by radioactivity. Although the amount of $^{56}$Ni
 derived from the tail luminosity and peak luminosity are consistent with each
 other, the secondary peak in the light curve makes this scenario
 unlikely. The data during the late phases does not allow us to
 constrain the onset of the optically thin phase powered only by radioactive
 decay. If we consider that intrinsically the \sn\ light curve is symmetric
 around the radioactively powered peak while the slow decline is instead
 produced by CSM interaction, we can replace $\tau_m$ by
 $\tau_{\rm ris}=34$\,d in the two above expressions and get
 $M_{\rm ej}\approx 14$\,\msun\ and $M_{\rm Ni}\approx 1.3$\,\msun.
\begin{figure}
\centering
\includegraphics[width=8.5cm]{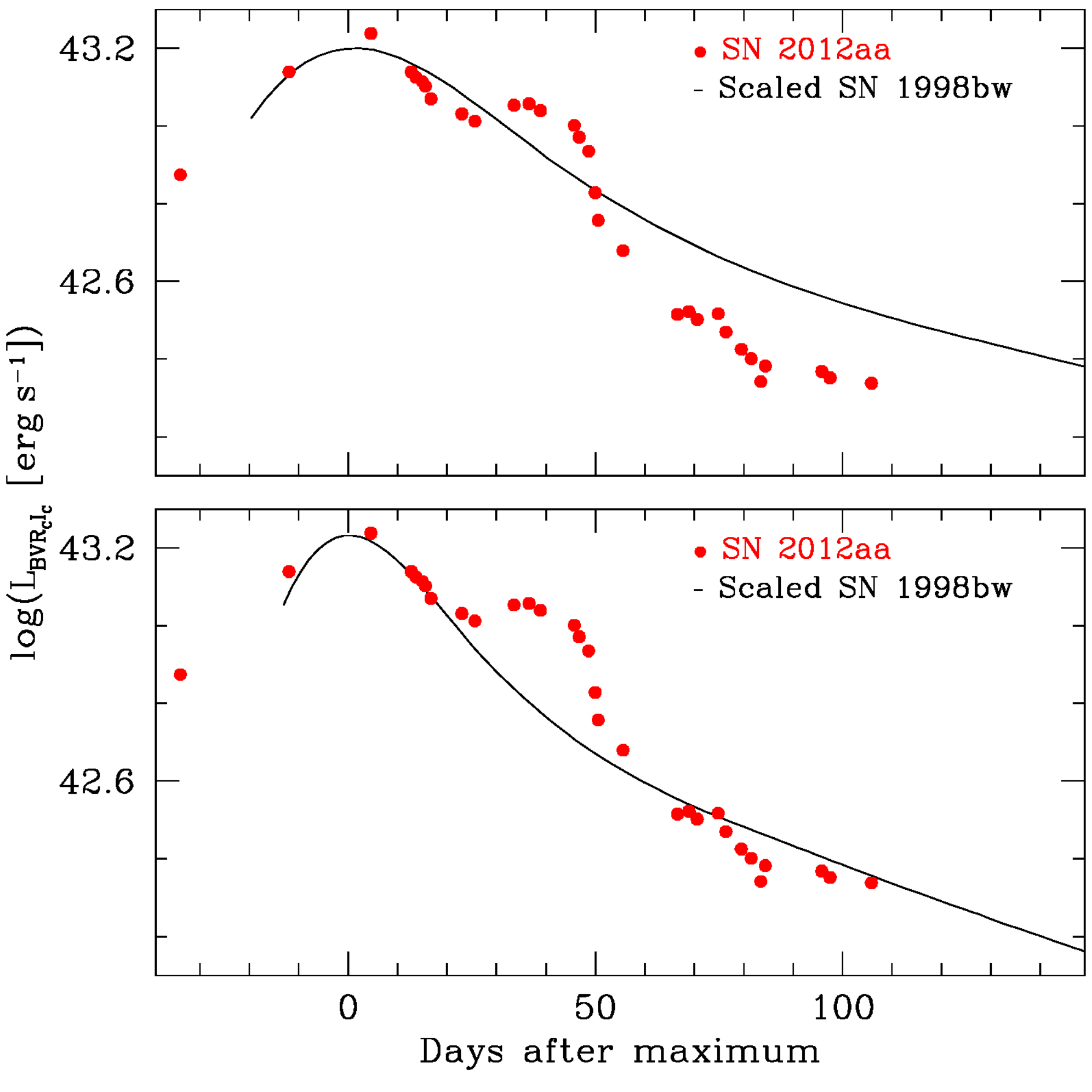}%
\caption{The \bvri\ quasi-bolometric light curve of SN 1998bw is fitted to that
 of \sn\ by stretching and scaling. {\bf Lower panel:} for this
 fit the data points of \sn\ during its secondary bump have not been considered.
 {\bf Upper panel:} for this fit the data points during the secondary bump have
 been considered. See text for details.}
\label{fig:cano} 
\end{figure}

 In another approach \citep{2013MNRAS.434.1098C}, we also use the \bvri\
 quasi-bolometric light curve of SN 1998bw published by
 \citet{2011AJ....141..163C} as
 a template and fit it to that of \sn\ by stretching along the time axis and
 scaling in luminosity; see Fig. \ref{fig:cano}. In
 the lower panel we have not considered the data points of \sn\
 observed during the secondary peak, giving 
 $M_{\rm ej}=10.9$--11.7\,\msun\ and
 $M_{\rm Ni}=1.3$--1.4\,\msun. These results are consistent with the values
 obtained from Arnett's rule considering only $\tau_{\rm ris}$. However, as
 shown in the Fig. \ref{fig:cano}, this process cannot fit the very first
 observation.
 We also fit the peak of the light curve considering also the bump (upper
 panel, Fig. \ref{fig:cano}). Obviously, it will not fit the late part, but
 it can well describe the rise. The obtained values for this fit are
 $M_{\rm ej}=28.4$--30.0\,\msun\ and $M_{\rm Ni}=1.2$--1.3\,\msun. These values
 are reasonably consistent  with results obtained by considering the
 higher-velocity profile of \Feii\ lines with peak
 $v_{\rm phot}=18,000$\,\kms\ and $\tau_{\rm ris}=34$\,d:
 $M_{\rm ej}=22.9$\,\msun\ and $M_{\rm Ni}=1.3$\,\msun.

 For the rest of this work, after comparing the spectroscopic and photometric
 properties of \sn\ with those of other CCSNe and SLSNe, we have adopted
 the following approximate values for the ejected mass and amount of $^{56}$Ni
 produced by \sn\,:
 $M_{\rm ej}\gtrsim14$\,\msun\ and $M_{\rm Ni}\approx 1.3$\,\msun.
 Hence, the kinetic energy of the explosion \citep{1996snai.book.....A} is
 roughly $E_K = 0.3\,M_{\rm ej}\,{v_{\rm phot}}^2 \approx 5.4\times10^{51}$
 ergs.

\begin{figure}
\centering
\includegraphics[width=8.5cm]{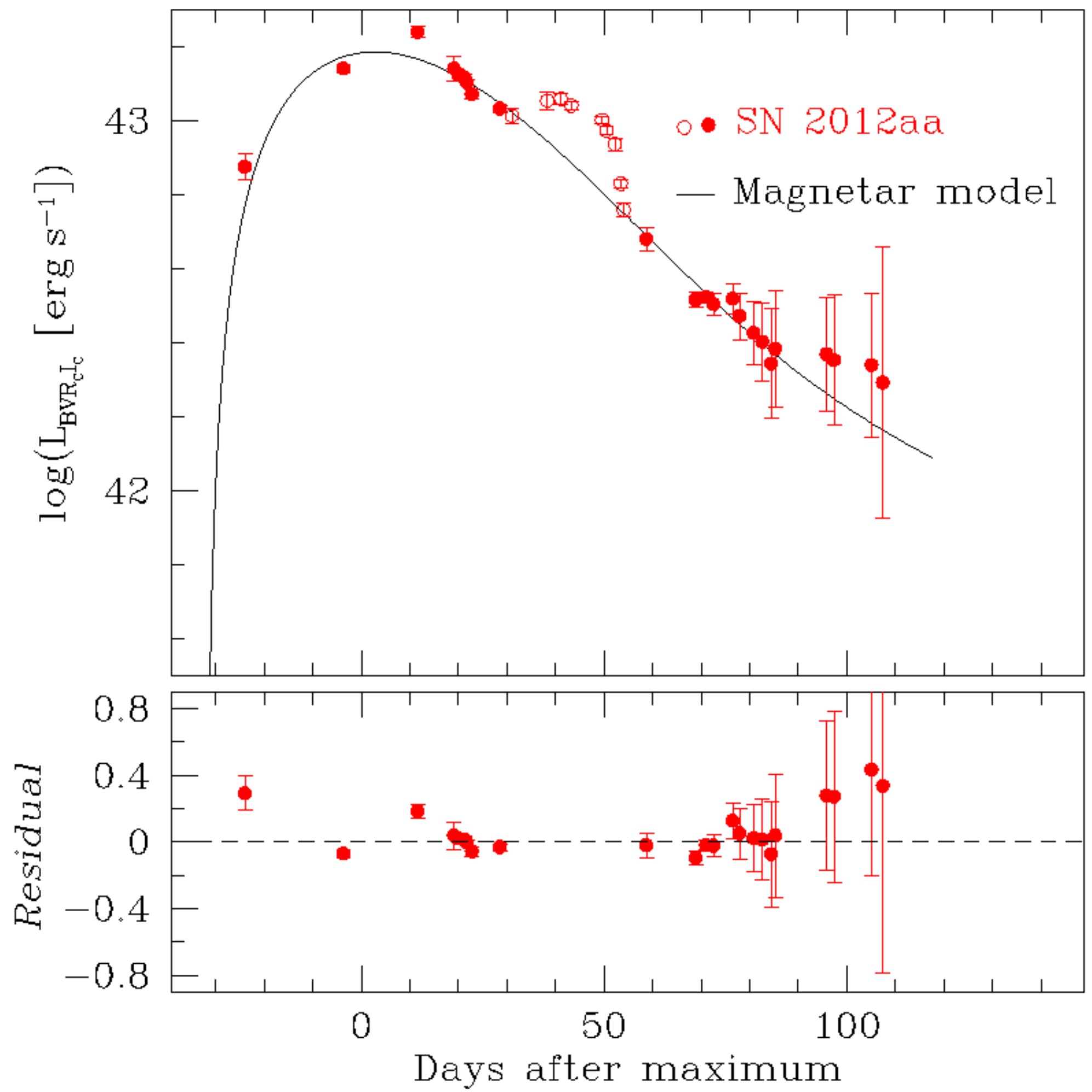}%
\caption{The magnetar model is fitted to the observed \bvri\
 quasi-bolometric light curve of \sn. We adopted the model proposed by
 \citet{2013ApJ...770..128I}. However, we only consider the power supplied
 by the magnetar, not that of radioactive decay. The errors associated with the
 observations have been calculated after propageting the errors in magnitudes
 only.
 While getting the best fit model we have excluded the data that corresponds to
 the bump in the lightcurve (marked by the open circles). The $Residual$ of the
 best fitted model has been shown in the lower panel along with the fitted data
 points (filled circles). Here $Residual$ is defined as:
 $Residual = (Observed data - Model data)/Model data$.
 For the best-fit model we found $M_{\rm ej} = 16.0\pm1.7$\,\msun,
 the magnetic field of the proto-magnetar $B = (6.5\pm0.3)\times10^{14}$\,G, 
 and the initial spin period $P = 7.1\pm0.6$\,ms.}
\label{fig:magnetar} 
\end{figure}

\subsection{CSM interaction powering} \label{res:env}
 The presence of a secondary bump in the light curves implies that radioactivity
 is not the only power source for \sn. Another way to
 get such high luminosity is CSM interaction. For strong CSM interaction,
 a dense environment is required. Comparing the rise and decline timescales of
 \sn\ with those of other interacting SLSNe and SNe~Ic
 (\citealt{2015MNRAS.452.3869N}, and references therein), suggests that
 $\sim$5 -- 10\,\msun\ of CSM was expelled by the progenitor,
 while the amount of ejected mass could be $\sim$ 10 -- 15\,\msun. 

 We compare ourlight curves with with results from recent
 hydrodynamical simulations from
 \citet{2015arXiv151000834S}, where the importance of CSM interaction in
 SLSNe was investigated. According to this study, an optical light curve
 with a sharp rise ($\sim20$\,d) before peak along with a fast decline after
 +50\,d is possible in an explosion with ejected mass of only 0.2\,\msun\ and
 kinetic energy $\sim2\times10^{51}$\,ergs within a 9.7\,\msun\ CSM having a
 wind profile with index $\sim1.8$.
 On the other hand, a longer rise time ($\sim50$\,d) and shallow decline in the
 optical post-peak light curve was achieved by an explosion with ejected
 mass 5\,\msun\ and kinetic energy $\sim4\times10^{51}$\,ergs within a 
 49\,\msun\ CSM. 

 The light-curve properties of \sn\ fall in-between these two
 categories. Fine tuning of the parameters could likely get a better
 match for the pure CSM interaction scenario, but we also note that the
 simulations of \citet{2015arXiv151000834S} do not include the effects of
 $^{56}$Ni, which could be an oversimplification, and that the  existence of a
 secondary bump requires the CSM to be distributed in a dense shell rather than
 in a constant wind. \sn\ could be an example where a combination of both
 radioactive decay and CSM interaction is responsible for the light-curve
 evolution.

\subsection{Magnetar powering} \label{res:magnetar}
 A third scenario is the presence of a central engine --- a proto-magnetar
 releasing spin energy that eventually powers the light curve over a long
 time. Several SLSN bolometric light curves have been modeled using this
 prescription. Here we have adopted the semi-analytical model developed by
 \citet{2013ApJ...770..128I}. We assumed that the secondary bump in \sn\ is
 produced by some other physical process; hence, when modeling we do not try to
 fit the data obtained during that period.We assumed that the
 photospheric velocity during the principal peak is $\sim 12,000$\,\kms, the
 kinetic energy of the explosion is $\sim 10^{51}$\,ergs, and the rise time 
 is 30--35\,d.
 The best fit model obtained by following Levenberg-Marquardt (LM) least-squares
 minimization technique \citep{More1978} is shown in the upper panel of Fig.
 \ref{fig:magnetar}, while the
 lower panel shows the residualr. It gives the following values
 for the explosion parameters: 
 $M_{\rm ej} = 16.0\pm1.7$\,\msun, magnetic field of the proto-magnetar 
 $B = (6.5\pm0.3)\times10^{14}$\,G, and initial spin period $P =
 7.1\pm0.6$\,ms. Apart from the bump, the model
 describes the overall broad observed light curve of \sn; The model gets
 slightly deviated (however within error) from the observed light curve
 beyond +60\,d past maximum.

\section{Discussion in the context of other events} \label{res:comp}
 From the above analysis it seems that \sn\ is a intermediate object between
 normal SNe~Ic and SLSNe~I.
 In this section we discuss the optical characteristics of \sn\ in the
 context of SNe~Ic and SLSNe.

\subsection{Comparison of photometric characteristics}\label{res:comp.phot}
 The $R_c$-band absolute magnitude at peak ($M_R\approx-20$\,mag) of \sn\ is on
 the bright end, though consistent with, that of the SN~Ic
 distribution ($-19.0\pm1.1$\,mag; \citealt{2011ApJ...741...97D}).
 However, its broader peak and higher tail luminosity 
 require a larger amount of radioactive $^{56}$Ni and more ejected mass, if
 modeled with the Arnett law. In addition, the presence of the secondary
 bump in the light curve requires some additional physical process.

 As shown in Fig. \ref{fig:bol}, the peak bolometric luminosity of \sn\ is
 $\sim1.6\times10^{43}$\,\ergs\ ($M_{BVR_cI_c}\approx-19.9$\,mag), 
 which is higher
 than that of engine-driven explosions such as the Type Ic SNe 1998bw
 ($\sim7.5\times10^{42}$\,\ergs), 2006aj ($\sim5.9\times10^{42}$\,\ergs),
 2003jd ($\sim6.5\times10^{42}$\,\ergs), and the normal Type Ib SN 2007uy
 ($\sim6.4\times10^{42}$\,\ergs). However, it is much less than the
 average peak luminosity ($\sim8.2\times10^{43}$\,\ergs\ or $M_{\rm Bol}
 \approx-21.7$\,mag) of Type I SLSNe 
  \citep{2013MNRAS.431..912Q} as well as slowly declining
 SLSNe exemplified by PTF12dam \citep{2013Natur.502..346N} in Fig.
 \ref{fig:bol}. 
 In SLSN-I the typical decline rates after maximum brightness are much faster
 than the radioactive decay rate of $^{56}$Co. For example, 
 in SN 2010gx the $g$, $r$, and $i$
 decline rates were $\gamma_{g} \approx 0.135$, $\gamma_{r} \approx
 0.084$, and $\gamma_{i} \approx 0.086$\,mag\,d$^{-1}$.
 Similar measurements for SNe~Ibn like SN 2011hw give:
 $\gamma_{B} \approx 0.058$, $\gamma_{V} \approx 0.052$, $\gamma_{R}
 \approx 0.055$, and
 $\gamma_{I} \approx 0.047$\,mag\,d$^{-1}$. 
 The steeper decline than radioactive
 decay in these SNe was argued to be either due to powering by CSM interaction 
 with no $^{56}$Ni or caused by substantial dust formation at late epochs
 \citep{2010ApJ...724L..16P, 2015MNRAS.449.1921P}. On the other hand, in the
 slowly declining SLSNe (e.g., PTF12dam) or SLSN-R (e.g., SN 2007bi) the decline
 rate is much slower: for PTF12dam the values are $\gamma_{g} \approx 0.02$,
 $\gamma_{r} \approx 0.014$, and $\gamma_{i} \approx 0.015$\,mag\,d$^{-1}$, 
 while for SN
 2007bi these are $\gamma_{B} \approx 0.024$, $\gamma_{V} \approx 0.018$, 
 $\gamma_{R} \approx 0.012$, and $\gamma_{I} \approx 0.006$\,mag\,d$^{-1}$. 
 These values are comparable (but not identical) to $^{56}$Co decay.

 By studying the evolution of bolometric light curves of SLSNe,
 \citet{2015MNRAS.452.3869N}
 found that for H-poor SLSNe-I, $20 \lesssim \tau_{\rm dec} \lesssim 40$\,d
 (i.e., 0.05 $\textgreater R_{\rm dec} \textgreater$ 0.03\,mag\,d$^{-1}$), 
 while for SLSN-R $\tau_{\rm dec} \sim$\,55--90\,d (i.e., 0.02
 $\textgreater R_{\rm dec} \textgreater$ 0.01\,mag\,d$^{-1}$). The rise
 timescales of these two sets of
 SNe are respectively $10 \lesssim \tau_{\rm ris} \lesssim 25$\,d and 
 $25 \lesssim \tau_{\rm ris} \lesssim 40$\,d. 
 They also found that for the entire sample,
 $\tau_{\rm dec} \textgreater \tau_{\rm ris}$ and generally $\tau_{\rm dec}$ 
 increases with $\tau_{\rm ris}$, implying two possible distributions of 
 SLSNe in the $\tau_{\rm ris}$ vs. $\tau_{\rm dec}$ plane, where one set 
 of SLSNe (with smaller values of $\tau_{\rm ris}$ and $\tau_{\rm dec}$) 
 falls near the position of normal CCSNe.
\begin{figure}
\centering
\includegraphics[width=8.0cm,angle=0]{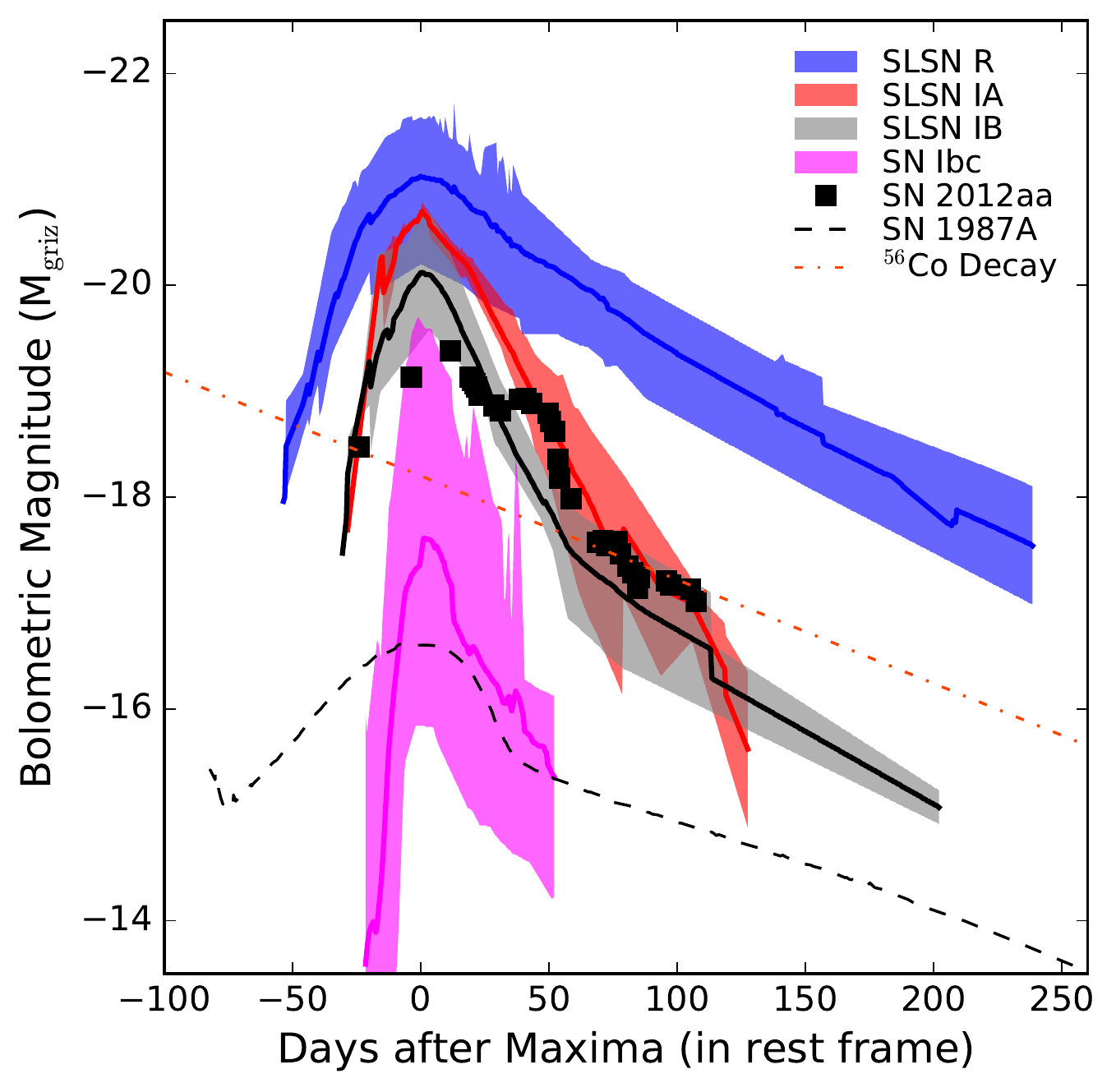}%
\caption{Comparison of \sn\ quasi-bolometric light curve with those 
 of SLSNe~Ic and SNe~Ibc. The SLSN datasets are obtained from
 \citet{2013ApJ...770..128I} and \citet{2015MNRAS.452.3869N}, while the
 SN~Ibc datasets are from \citet{2015A&A...574A..60T}. SLSN-R are
 distinctly different from SLSN-I. There may be two
 categories (A and B) of SLSN-I. The solid lines represent the mean of each
 population. The plot also presents the decay rate of $^{56}$Co and the 
 well-observed Type II SN 1987A, which had a nebular decline consistent 
 with $^{56}$Co decay.}
\label{fig:comp}
\end{figure}

 The rise and decline timescales of the bolometric light curve of \sn\ are
 $\tau_{\rm ris}=34$\,d and $\tau_{\rm dec}=57$\,d, which implies that
 this SN is an event in between the two aforementioned distributions.
 Its decline ($\sim0.03$\,mag\,d$^{-1}$) before the secondary bump is
 similar to that of SLSN-I and SNe~Ic, while the relatively slower decline
 (0.02\,mag\,d$^{-1}$) after
 +70\,d is similar to SLSN-R and not very different from that expected from
 $^{56}$Co decay. We also note that among the SLSNe-I (which behave similar to
 SNe~Ic in the $\tau_{\rm ris}$ vs. $\tau_{\rm dec}$ plane), some 
 events (e.g., SNe 2011ke, 2012il, PTF10hgi, LSQ14mo; see 
 \citealt{2015MNRAS.452.3869N}, their fig. 1) show a decline trend after
 $\sim60$\,d past maximum that is similar to \sn\ (i.e., 0.02\,mag\,d$^{-1}$)
 and comparable to $^{56}$Co decay.

 Figure \ref{fig:comp} presents the light curves of different SNe~Ibc and
 SLSNe~I along with that of \sn\footnote{The conversion of the 
quasi-bolometric luminosity of \sn\ to $M_{griz}$ magnitude was done 
 by following the relation \citep{2015MNRAS.452.3869N}
 \[M_{griz}=-2.5\,{\rm log}_{10}[L_{\rm Bol}/(3.055\times10^{35})].\]} 
 as well as the Type II SN 1987A which has a late time decline consistent with
 $^{56}$Co decay. Some SLSNe~I start to show flattening in their
 light curves at earlier phases than others in this group. Before +60\,d, they
 decline faster ($\gtrsim 0.03$\,mag\,d$^{-1}$), while at later stages the slope
 is more shallow ($\sim0.02$\,mag\,d$^{-1}$). It seems that the events PTF10hgi,
 SN 2011ke, SN 2012aa, and SN 2012il belong to a subset that
 exhibits the photometric characteristics of both SLSNe-I (potentially shock
 dominated) and SLSNe-R (potentially radioactivity dominated). 

 Based on Fig. \ref{fig:comp}, we tentatively
 divide SLSN~I into two categories (hereafter A and B). The SNe that fall inside the first set
 (category A) decline faster than the decay rate of $^{56}$Co, while the events
 that fall in the second set (category B) may show a radioactive decay phase
 ($\gtrsim60$\,d after maximum light). At a considerable time after maximum
 (e.g., at +150\,d as shown in figure), the second set becomes brighter than
 the first set, although dimmer than SLSNe-R. We also note that the mean peak
 width (measurement of the diffusion timescale) of category B is narrower than 
 for category A. More extensive follow-up observations of these events at late
 phases are required to further understand any different subclasses of these
 stellar explosions. 

\subsection{Comparison of host properties}\label{res:comp.gal}
 Using a sample of 23 SLSN hosts, \citet{2015MNRAS.449..917L}
 found that H-poor SLSNe mostly occur in extreme emission-line galaxies (EELGs;
 \citealt{2015A&A...578A.105A}), whereas SLSNe-II can be found in more massive,
 metal-rich, and relatively evolved galaxies. 
 In contrast to H-poor SLSNe, broad-lined SNe~Ic have not been found in 
 EELGs and are associated with evolved star-forming galaxies
 \citep{2008AJ....135.1136M}. In this context the host of \sn\ provides an
 example, with $M_B\approx-18.13$\,mag and $Z_{\rm host} = 0.92\pm0.34$\,\zsun.
 This makes it in between
 the ordinary star-forming galaxies and the hosts of GRB-associated events,
 where the latter preferentially occur in low-luminosity and low-metallicity
 environments \citep{2008AJ....135.1136M}.

 With an extended galaxy sample, \citet{2013A&A...550A..69A} found that the
 hosts in which only a single CCSN has been found have a median absolute
 magnitude as low as $-20.64$ in $B$ (also see
 \citealt{2013A&A...558A.143T} for the cumulative distribution of the $B$
 magnitudes of the host galaxies of CCSNe). The host of \sn\ has absolute $B$
 magnitude $\sim-18.13$. This implies that \sn\ occurred in a relatively faint
 galaxy compared to galaxies that normally host CCSNe.  

\subsection{Investigating other possibilities}\label{res:otherpossibility}
 There are other scenarios which may produce a light curve and/or
 spectra similar to those observed in \sn.

 As discussed in Sect. \ref{intro}, TDEs also fill the luminosity and
 timescale
 gap between CCSNe and SLSNe. However, these events are generally 
 expected to occur
 in galactic nuclei where the supermassive black hole (SMBH) resides.
 They could potentially be found at a location away from the center, 
 if the entire system has two SMBHs. \sn\ exploded in the outskirts 
 of its host, and in the
 late-time template image taken with the TNG we do not see any additional
 flux enhancement at the SN location. There is thius no evidence for a
 SMBH at the location of the SN. The spectral behaviour of \sn\ is also not
 comparable with those of known TDEs.

 If the emission near \ha\ is indeed due to H, it could be
 preferable to call \sn\ a Type II event. Several SNe~IIn and Ibn
 have been discovered with peak magnitudes between $\sim -19$ and $-20$
 (\citealt{2012AJ....144..131Z,2016MNRAS.456..853P},
 and references therein). In all such cases, strong, narrow emission features
 (from unshocked CSM) with a broad component (from shocked CSM, predominantly H
 and/or He) have been observed. This was not seen in \sn\,. The
 impression of a narrow \ha\ line in the first three spectra is
 instead most
 likely caused by contamination by emission from the host center, which was
 unresolved from the SN during the observations. 
 Spectral comparison with other events shows that \sn\ is of Type~Ic. 
 The colour evolution of \sn\ is also more similar to that of SNe~Ibc
 rather than to SNe~IIn, where a monotonic increase of colour
 with time is usually observed \citep{2012AJ....144..131Z}.

 As discussed in Sect. \ref{intro}, the Type
 Ibc SN 2005bf and a few SLSNe~I also showed double peaks in their light curves.
 In contrast, \sn\ exhibited a short-duration peak after the principal broader
 peak. In the ``relativistic jet'' or ``high-pressure bubble''
 explosion scenarios, the short-duration bump is expected to occur before the
 broader peak (because of the shorter dynamical timescale).  We can
 therefore rule out
 these two mechanisms for the formation of the secondary bump in
 \sn. Nevertheless, the presence of a spin-down magnetar as a central source
 of energy cannot be ruled out; the broader peak of \sn\ could be 
 powered by a spin-down magnetar
 However, to explain the secondary bump in \sn, an
 excess supply of energy from the central source would be required. An
 asymmetrically ejected $^{56}$Ni blob
 might also explain the light curve under some special ejecta geometry, but
 detailed modeling would be required to explore such a scenario.  

 The spectroscopic similarity of SNe 2003jd, 2006aj, and 2012aa (left panel
 of Fig. \ref{fig:spcmp}) makes it important to discuss the possible GRB/XRF
 connection with \sn. SN 2006aj was the counterpart of XRF 060218
 \citep{2006GCN..4775....1C}, while SN 2003jd may have had a GRB connection,
 although a GRB was not detected \citep{2008MNRAS.383.1485V}. So far, no direct
 evidence of a GRB/XRF association with H-poor SLSNe has been reported; although
 the comparable host metallicities of these two types of events have been 
 noticed \citep{2014ApJ...787..138L, 2015MNRAS.449..917L}.
 We have also searched the archive of the Interplanetary Network
 (IPN)\footnote{http://heasarc.gsfc.nasa.gov/w3browse/all/ipngrb.html} for
 the possible detection of any GRB/XRF in the direction of \sn\,. However no
 burst was detected in that direction. 
 Thus, probably there is little or no GRB connection with \sn.


\section{Conclusions} \label{concl}
 The nature of the progenitors and the explosion mechanism of SLSNe is still 
 unclear. Whether there is a difference in the explosion mechanisms of SLSNe
 and canonical SNe~Ibc is also unknown. In this context, the
 study of luminous events that fall between CCSNe and SLSNe in various respects
 is important. Here we have carried out optical photometric and spectroscopic
 observations of \sn\ over an period of 120\,d.
 In this investigation, we proposed that \sn\ is a Type~Ic SN that 
 occurred in a dense medium, which made the event more luminous owing to
 SN-CSM interaction and resulting in a photometric evolution similar to that
 of H-poor SLSNe. Different aspects of this study include the following.

{The peak bolometric luminosity of \sn\ is
 $\sim1.6\times10^{43}$\,ergs ($M_{\rm Bol}\approx-20$\,mag), 
 which is less than the
 typical peak luminosity of SLSNe~I but larger than that of SNe~Ic. 
 The spectroscopic properties of \sn\ are similar to those of normal
 SNe~Ic. However, its rise timescale ($\tau_{\rm ris}=34$\,d) and decline
 timescale ($\tau_{\rm dec}=57$\,d) are more consistent with those of 
  H-poor SLSNe.}

 {The event shows a secondary bump in all optical bands between
 +40\,d and +55\,d after maximum brightness. 
 It is more pronounced in the $R_c$ and $I_c$ bands
 than in the $B$ and $V$ bands.}

 {Beyond +47\,d after maximum brightness, 
 we noticed an emission peak in
 the spectra at rest wavelength 6500\,\AA. There are two possibilities:
 either this is \Oia\ $\lambda\lambda$6300, 6364 or \ha\ in
 emission. In the first scenario, the line-emitting region is moving away with a
 projected velocity of $\sim9500$\,\kms.
 In the second scenario the emission is 
 \ha\ blueshifted by 3000\,\kms;
 since it appeared at late epochs, it is probably part of the CSM and was
 powrwed by interaction. 
 The identification of \ha\ is consistent with other line profiles.
 On the other hand, comparison of spectra of \sn\ with those of other
 SNe~Ic and SLSNe suggests that this particular feature is more likely
 \Oia\ rather than \ha.}

 {Assuming that only $^{56}$Ni decay is responsible for powering
 the SN, from the quasi-bolometric light curve 
 we require roughly 1.3\,\msun\
 of $^{56}$Ni ejected in this explosion. The ejected mass is 
 $\sim 14$\,\msun, implying a kinetic energy of 
 $\sim5.4\times10^{51}$\,erg. On the
 other hand, if the entire explosion is CSM-interaction dominated, a similar
 explosion can be produced under the presence of CSM with a mass of
 $\sim5$--10\,\msun.}

 {We have also explored the possibility of the emergence of a
 magnetar. Assuming that the velocity of the ejecta near peak luminosity is 
 $\sim12,000$\,\kms, the kinetic energy of the explosion is 
 $\sim10^{51}$\,ergs, and the rise time is 30--35\,d, we find that 
 the quasi-bolometric light curve of \sn\ can be fitted with a 
 magnetar having a magnetic field of $(6.5\pm0.3)\times10^{14}$\,G, an
 initial spin period of $7.1\pm0.6$\,ms, and an explosion ejected mass 
 of $16.0\pm1.7$\,\msun.}

 {With a limited SN sample consisting of PTF10hgi, 
 SN 2011ke, SN 2012aa, and SN 2012il, 
 we found that there is a potential subset of
 SLSNe~I showing a radioactivity-powered tail at relatively early
 times ($\sim 60$\,d after maximum brightness). 
  This subset also has a narrower peak (i.e., smaller 
 diffusion time) in comparison with other SLSN~I/SLSN-R events.}

 {The $z\approx0.08$ host of \sn\ is a
 star-forming (Sa/Sb/Sbc) galaxy.
 The overall metallicity ($Z_{\rm host}\approx0.92\pm0.34$\,\zsun) 
 of the host is comparable to the solar metallicity and also to 
 the metallicities of typical nearby spiral galaxies hosting SNe~IIP and
 normal SNe~Ibc.}

\begin{acknowledgements}
 We thank all of the observers at ST who provided their valuable time and
 support for observations of \sn. We are grateful to the staff of the 
 CT, IGO, TNG, NTT, Lick, and Keck-I telescopes for their kind cooperation 
 when conducting the observations. This work is partially based on
 observations made with the NOT, operated by the Nordic Optical Telescope
 Scientific Association at the Observatorio del Roque de los Muchachos, La
 Palma, Spain, of the Instituto de Astrofisica de Canarias; 
 on observations made with the Italian
 Telescopio Nazionale Galileo (TNG) operated on the island of La Palma by the
 Fundación Galileo Galilei of the INAF (Istituto Nazionale di Astrofisica) at
 the Spanish Observatorio del Roque de los Muchachos of the Instituto de
 Astrofisica de Canarias; and on observations made with NTT
 Telescope at the La Silla and Paranal Observatories within the European
 supernova collaboration involved in ESO-NTT large programme 184.D-1140 led by
 Stefano Benetti. Some of the data presented herein were obtained at the W. M.
 Keck Observatory, which is operated as a scientific partnership among the
 California Institute of Technology, the University of California, and the
 National Aeronautics and Space Administration (NASA); the observatory was made
 possible by the generous financial support of the W. M. Keck Foundation. We
 thank G. Leloudas for a valuable discussion of a draft of this paper.  J.M.S.
 is supported by an NSF Astronomy and Astrophysics Postdoctoral Fellowship under
 award AST-1302771. A.P., E.C., S.B., and L.T. are partially supported by the
 PRIN-INAF 2014 under the project ``Transient Universe: unveiling new types of
 stellar explosions with PESSTO.''A.V.F.'s research is supported by the
 Christopher R. Redlich Fund, the TABASGO Foundation, and NSF grant AST-1211916.
 Research at Lick Observatory is partially supported by a generous gift from
 Google. This work has made use of the NASA/IPAC Extragalactic Database
 (NED), which is operated by the Jet Propulsion Laboratory, California Institute
 of Technology, under contract with NASA. We also used NASA's Astrophysics Data
 System.
\end{acknowledgements}


\newpage
\begin{table*}
\setcounter{table}{3}
\centering
\caption{CRTS photometry of SN 2012aa$^a$.\label{tab:crtslog}}
\vskip 0.2cm
\scriptsize
\begin{tabular}{cccccc}
\hline\hline
UT Date&JD $-$ &Phase$^b$&$V$&$V_{\rm err}$&Seeing$^c$\\
(yyyy/mm/dd)& 2,450,000& (day)& (mag)& (mag)& (\arcsec)\\
\hline
2004/1/27.53 & 3032.03 & -2919.97 & 18.94 & 0.20 & 2.46	\\
2004/4/13.90 & 3108.90 & -2843.10 & 18.81 & 0.14 & 2.56	\\
2004/4/30.87 & 3125.87 & -2826.13 & 18.86 & 0.19 & 1.88	\\
2004/5/1.83 & 3126.83 & -2825.17 & 18.87 & 0.14 & 3.04	\\
2004/6/6.77 & 3162.77 & -2789.23 & 19.00 & 0.22 & 4.00	\\
2004/6/19.77 & 3175.77 & -2776.23 & 18.98 & 0.08 & 2.58	\\
2004/6/25.71 & 3181.71 & -2770.29 & 18.90 & 0.15 & 2.76	\\
2005/1/18.50 & 3389.00 & -2563.00 & 19.13 & 0.05 & 2.91	\\
2005/3/4.95 & 3433.95 & -2518.05 & 19.10 & 0.10 & 2.96	\\
2005/4/5.89 & 3465.89 & -2486.11 & 19.17 & 0.06 & 3.76	\\
2005/4/18.86 & 3478.86 & -2473.14 & 19.22 & 0.13 & 3.83	\\
2005/5/6.79 & 3496.79 & -2455.21 & 19.02 & 0.05 & 2.90	\\
2005/5/11.86 & 3501.86 & -2450.14 & 19.13 & 0.09 & 2.63	\\
2005/6/4.82 & 3525.82 & -2426.18 & 19.11 & 0.09 & 2.66	\\
2006/1/9.52 & 3745.02 & -2206.98 & 19.15 & 0.06 & 2.80	\\
2006/1/26.51 & 3762.01 & -2189.99 & 19.12 & 0.10 & 3.09	\\
2006/2/25.93 & 3791.93 & -2160.07 & 19.06 & 0.12 & 3.48	\\
2006/4/19.92 & 3844.92 & -2107.08 & 18.99 & 0.18 & 2.62	\\
2006/4/30.85 & 3855.85 & -2096.15 & 19.10 & 0.10 & 2.73	\\
2006/5/7.78 & 3862.78 & -2089.22 & 19.08 & 0.19 & 2.90	\\
2006/5/20.82 & 3875.82 & -2076.18 & 19.06 & 0.07 & 2.93	\\
2006/5/26.85 & 3881.85 & -2070.15 & 19.16 & 0.06 & 2.70	\\
2006/6/4.85 & 3890.85 & -2061.15 & 19.04 & 0.35 & 2.71	\\
2006/7/21.72 & 3937.72 & -2014.28 & 19.15 & 0.20 & 3.05	\\
2007/1/27.51 & 4128.01 & -1823.99 & 19.16 & 0.17 & 3.17	\\
2007/2/23.50 & 4155.00 & -1797.00 & 19.09 & 0.07 & 2.77	\\
2007/3/16.94 & 4175.94 & -1776.06 & 19.16 & 0.10 & 2.87	\\
2007/3/25.93 & 4184.93 & -1767.07 & 19.18 & 0.07 & 2.70	\\
2007/4/15.88 & 4205.88 & -1746.12 & 19.15 & 0.06 & 2.57	\\
2007/4/22.78 & 4212.78 & -1739.22 & 19.11 & 0.12 & 2.90	\\
2007/5/9.89 & 4229.89 & -1722.11 & 19.13 & 0.07 & 2.87	\\
2007/5/17.82 & 4237.82 & -1714.18 & 19.05 & 0.18 & 2.99	\\
2007/5/24.81 & 4244.81 & -1707.19 & 19.19 & 0.08 & 2.77	\\
2007/6/8.76 & 4259.76 & -1692.24 & 19.10 & 0.10 & 2.60	\\
2007/6/16.72 & 4267.72 & -1684.28 & 19.14 & 0.16 & 2.72	\\
2008/1/11.50 & 4477.00 & -1475.00 & 19.10 & 0.04 & 2.86	\\
2008/2/7.51 & 4504.01 & -1447.99 & 19.10 & 0.07 & 3.25	\\
2008/2/13.99 & 4509.99 & -1442.01 & 19.14 & 0.12 & 2.55	\\
2008/2/28.52 & 4525.02 & -1426.98 & 19.00 & 0.12 & 2.59	\\
2008/3/5.86 & 4530.86 & -1421.14 & 19.05 & 0.10 & 2.63	\\
2008/3/29.93 & 4554.93 & -1397.07 & 19.14 & 0.16 & 2.93	\\
2008/4/11.88 & 4567.88 & -1384.12 & 19.18 & 0.15 & 2.78	\\
2008/5/11.87 & 4597.87 & -1354.13 & 19.13 & 0.10 & 2.59	\\
2008/12/30.54 & 4831.04 & -1120.96 & 19.13 & 0.11 & 2.98\\
2009/1/31.50 & 4863.00 & -1089.00 & 19.09 & 0.08 & 3.02	\\
2009/3/19.97 & 4909.97 & -1042.03 & 19.13 & 0.01 & 3.42	\\
2009/4/28.90 & 4949.90 & -1002.10 & 19.14 & 0.08 & 2.70	\\
2009/5/13.78 & 4964.78 & -987.22 & 19.03 & 0.01 & 4.13	\\
2009/5/24.79 & 4975.79 & -976.21 & 19.17 & 0.04 & 2.76	\\
2009/6/13.73 & 4995.73 & -956.27 & 19.24 & 0.15 & 1.88	\\
2009/6/14.73 & 4996.73 & -955.27 & 19.14 & 0.17 & 3.11	\\
2010/2/15.98 & 5242.98 & -709.02 & 19.12 & 0.19 & 2.63	\\
2010/3/13.50 & 5269.00 & -683.00 & 19.14 & 0.06 & 2.61	\\
2010/3/21.97 & 5276.97 & -675.03 & 19.08 & 0.13 & 2.86	\\
2010/4/9.84 & 5295.84 & -656.16 & 19.02 & 0.12 & 4.48	\\
2010/5/5.87 & 5321.87 & -630.13 & 19.12 & 0.03 & 2.76	\\
2010/5/19.76 & 5335.76 & -616.24 & 19.07 & 0.17 & 3.27	\\
2010/6/2.77 & 5349.77 & -602.23 & 19.07 & 0.08 & 3.21	\\
2010/6/9.68 & 5356.68 & -595.32 & 19.06 & 0.13 & 2.74	\\
2010/6/17.71 & 5364.71 & -587.29 & 19.10 & 0.19 & 3.00	\\
2011/1/15.55 & 5577.05 & -374.95 & 18.99 & 0.12 & 2.98	\\
2011/1/30.52 & 5592.02 & -359.98 & 19.10 & 0.05 & 3.32	\\
2011/2/12.50 & 5605.00 & -347.00 & 19.13 & 0.11 & 2.93	\\
2011/3/1.50 & 5622.00 & -330.00 & 19.18 & 0.07 & 3.70	\\
2011/3/14.93 & 5634.93 & -317.07 & 19.07 & 0.10 & 3.23	\\
2011/3/27.90 & 5647.90 & -304.10 & 19.09 & 0.19 & 3.23	\\
2011/4/3.91 & 5654.91 & -297.09 & 19.19 & 0.20 & 1.88	\\
2011/4/13.88 & 5664.88 & -287.12 & 19.08 & 0.12 & 3.16	\\
2011/4/23.91 & 5674.91 & -277.09 & 18.99 & 0.22 & 2.77	\\
2011/5/6.84 & 5687.84 & -264.16 & 19.17 & 0.20 & 2.67	\\
2011/5/13.82 & 5694.82 & -257.18 & 19.22 & 0.03 & 2.82	\\
2011/5/20.80 & 5701.80 & -250.20 & 19.07 & 0.18 & 4.26	\\
2011/6/8.75 & 5720.75 & -231.25 & 19.20 & 0.19 & 2.75	\\
2011/6/22.70 & 5734.70 & -217.30 & 19.14 & 0.14 & 4.12	\\
2012/1/3.53 & 5930.03 & -21.97 & 18.41 & 0.07 & 3.35	\\
2012/1/25.52 & 5952.02 & 0.02 & 18.02 & 0.05 & 2.82	\\
2012/2/21.50 & 5979.00 & 27.00 & 18.11 & 0.05 & 2.55	\\
2012/2/29.98 & 5986.98 & 34.98 & 18.22 & 0.04 & 2.77	\\
2012/3/16.88 & 6002.88 & 50.88 & 18.21 & 0.07 & 2.66	\\
2012/3/27.96 & 6013.96 & 61.96 & 18.43 & 0.11 & 2.94	\\
2012/4/15.93 & 6032.93 & 80.93 & 18.77 & 0.18 & 2.98	\\
2012/4/21.85 & 6038.85 & 86.85 & 18.73 & 0.08 & 2.55	\\
\end{tabular}
\end{table*}
\newpage
\begin{table*}
\centering
Continued from previous page
\vskip 0.2cm
\scriptsize
\begin{tabular}{cccccc}
\hline\hline
UT Date&JD $-$ &Phase$^b$&$V$&$V_{err}$&Seeing$^c$\\
(yyyy/mm/dd)& 2,450,000 & (day)& (mag)& (mag)& (\arcsec)\\
\hline
2012/5/12.82 & 6059.82 & 107.82 & 18.91 & 0.10 & 2.79	\\
2012/5/22.85 & 6069.85 & 117.85 & 18.94 & 0.13 & 3.34	\\
2012/6/9.78 & 6087.78 & 135.78 & 18.90 & 0.05 & 3.30	\\
2012/6/15.73 & 6093.73 & 141.73 & 18.97 & 0.08 & 2.77	\\
2013/1/12.53 & 6305.03 & 353.03 & 19.03 & 0.22 & 2.74	\\
2013/3/14.96 & 6365.96 & 413.96 & 19.27 & 0.15 & 3.04	\\
2013/4/2.92 & 6384.92 & 432.92 & 19.15 & 0.24 & 2.69	\\
2013/4/11.88 & 6393.88 & 441.88 & 19.07 & 0.18 & 2.72	\\
2013/5/5.77 & 6417.77 & 465.77 & 19.19 & 0.06 & 2.76	\\
2013/5/16.83 & 6428.83 & 476.83 & 19.06 & 0.11 & 3.14	\\
2013/5/30.75 & 6442.75 & 490.75 & 19.17 & 0.03 & 2.79	\\
2013/6/6.77 & 6449.77 & 497.77 & 19.03 & 0.10 & 2.88	\\
2013/6/15.73 & 6458.73 & 506.73 & 19.15 & 0.14 & 3.04	\\
2013/6/27.73 & 6470.73 & 518.73 & 19.12 & 0.13 & 3.04	\\
\hline
\end{tabular}
\tablefoot{
   \tablefoottext{a}{All of the unfiltered CRTS magnitudes have been calibrated to
  the Johnson $V$-filter system using the local sequences.}\newline
   \tablefoottext{b}{The event was dicovered on JD 2,455,956.04. However, the
  $V$-band maximum was estimated to be about 4\,d prior to discovery. Here, 
 phases are calculated with respect to the epoch of $V$ maximum in the observed
  frame, corresponding to JD 2,455,952. For details, see Sect. \ref{phot}.}
 \newline
   \tablefoottext{c}{FWHM of the stellar PSF in the $V$ band.}\newline
}
\end{table*}
 
\end{document}